\documentclass[preprint,aps,pra,superscriptaddress,floatfix,nofootinbib]{revtex4-1}
 \usepackage{amsmath}
\usepackage{amssymb}
\usepackage{graphicx}
\usepackage{xfrac}
\usepackage[usenames,dvipsnames]{xcolor}
\usepackage{hyperref}
\hypersetup{
    plainpages=false,       
    unicode=false,          
    pdftoolbar=true,        
    pdfmenubar=true,        
    pdffitwindow=false,     
    pdfstartview={FitH},    
    pdftitle={Probing the Ocean's Multiscale Pathways},
    pdfauthor={Storer, Hadi-Sichani, and Aluie},
    pdfnewwindow=true,      
    colorlinks=true,        
    linkcolor=black,         
    citecolor=cyan,      
    filecolor=black,      
    urlcolor=olive           
}
\usepackage{subfigure}
\usepackage[normalem]{ulem}
\usepackage{natbib}

{\vskip 2pt \begin{compactitem}[#1]\setlength{\itemsep}{2pt}}
{\end{compactitem}\vskip 2pt}

\newcommand{\be}{\begin{equation}}
\newcommand{\ee}{\end{equation}} 
\newcommand{\lb}{\label}
\newcommand{\OL}{\overline}
\newcommand{\wh}{\widehat}

\newcommand{\md}{{\mathrm{d}}}
\newcommand{\vell}{{\vec{\ell}}}

\newcommand{\ba}{{\bf a}}

\newcommand{\bk}{{\bf k}}

\newcommand{\br}{{\bf r}}
\newcommand{\bu}{{\bf u}}
\newcommand{\bw}{{\bf w}}

\newcommand{\bx}{{\bf x}}

\newcommand{\bK}{{\bf K}}

\newcommand{\mE}{{\mathcal{E}}}

\newcommand{\bzed}{{\mbox{\boldmath $0$}}}

\begin{document}

\title{\textbf {Measuring Scale-dependent Shape Anisotropy by Coarse-Graining:\\ Application to Inhomogeneous Rayleigh-Taylor Turbulence}}
\author{Dongxiao Zhao}
\email{dzhao5@ur.rochester.edu}
\affiliation{Department of Mechanical Engineering, University of Rochester}
\affiliation{School of Naval Architecture, Ocean and Civil Engineering, Shanghai Jiao Tong University}
\author{Hussein Aluie}
\affiliation{Department of Mechanical Engineering, University of Rochester}
\affiliation{Laboratory for Laser Energetics, University of Rochester}

\begin{abstract}
    We generalize the `filtering spectrum' \citep{Sadek18} to probe scales along different directions by spatial coarse-graining. This multi-dimensional filtering spectrum quantifies the spectral content of flows that are not necessarily homogeneous. 
   From multi-dimensional spectral information, we propose a simple metric for shape anisotropy at various scales.
  The method is applied to simulations of 2D and 3D Rayleigh-Taylor (RT) turbulence, which is inhomogeneous and anisotropic. We show that 3D RT has clear shape anisotropy at large scales with approximately $4:3$ vertical to horizontal aspect ratio, but tends toward isotropy at small scales as expected \cite{Chertkov03,Soulard12,Zhou17-2}. In sharp contrast, we find that RT in 2D simulations, which are still the main modeling framework for many applications,  is isotropic at large scales and its shape anisotropy increases at smaller scales where structures tend to be horizontally elongated. While this may be surprising, it is consistent with recent results in \citep{Zhao22JFM}; large-scale isotropy in 2D RT is due to the generation of a large-scale overturning circulation via an upscale cascade, while small scale anisotropy is due to the stable stratification resultant from such overturning and the inefficient mixing in 2D.
\end{abstract}

\maketitle

\section{Introduction}
Flows encountered in nature or engineering are often anisotropic. Anisotropy can arise from the driving mechanisms, boundary conditions, or body forces that break rotational symmetry such as those due to gravity, rotation, or magnetic fields. Examples range from the motion of polymer fluids \cite{graham2011fluid,benzi2018polymers} and quantum superfluids \cite{Biferale19PRL,Yui20PRL}, to flows in geophysics \cite{Manning86AAP,Grechko92ASR,Toselli11OSA,Cui15OSA,Pouquet19POF} and astrophysics \cite{Shebalin83JPP,Montgomery95APJ,Cho00APJ,Horbury08PRL}, including free shear flows such as jets and plumes \cite{list1982turbulent}, and bounded flows such as in a channel \cite{Antonia91JFM,Rasam11JOT}.

In this work, we are concerned with \emph{shape anisotropy} and not with anisotropy due to the vector components of a flow's velocity. To distinguish the two notions of anisotropy, we shall call the latter \emph{vector anisotropy}. In the turbulence literature, these two notions of anisotropy are related to so-called \emph{directional} and \emph{polarization}  anisotropy, respectively \cite{Cambonetal97JFM,Cambon01EJMB,GodeferdChantal03JFM,cambon2006anisotropic,Burlotetal15POF}. We avoid this terminology since it is suggestive of wave phenomena. Here, we are motivated by more general flow structures that may be spatially localized.
Shape anisotropy and vector anisotropy are often correlated, although the former can arise in scalar fields such as density or temperature. Shape anisotropy of the velocity field may be diagnosed for each velocity component separately or by analyzing the flow's kinetic energy (KE) as we do below. 

In canonical turbulence that may be described by Kolmogorov's theory \cite{Kolmogorov41}, any anisotropy present at the large scales is expected to decrease at smaller scales at sufficiently high Reynolds numbers, where the flow is statistically isotropic \cite{Pope2001}. However, this is not always the case, such as in magnetohydrodynamic turbulence \cite{goldreich1997magnetohydrodynamic,Boldyrev05APJL,Bian19,schekochihin2022mhd} where anisotropy is expected to be more pronounced at smaller scales. For the objective testing of turbulence phenomenologies, and for understanding and modeling complex flows such as those mentioned above, the community has relied on a variety of metrics that quantify anisotropy at different scales.

The choice of anisotropy metrics depends on the scales of interest. Bulk turbulence (single-point) anisotropy is traditionally quantified using invariants of the Reynolds stress tensor \cite{Lumley77JFM,Livescu09,Zhou17-2,Pouquet19POF}.
At the smallest scales, anisotropy is often measured from the components of gradients \cite{Thiesset13JFM,Carter17JFM,Garanaik18POF}.
At intermediate (possibly inertial) scales, anisotropy is often characterized using two-point statistics such as autocorrelation functions, Fourier spectra, or 2nd-order structure functions \cite{yeung1991response,AradLvov99PRE,BiferaleProcaccia,KurienSreenivasan00PRE,BiferaleToschi01PRL,ishihara2002anisotropic,Shen02POF,cambon2006anisotropic,Casciola07POF,burlot2015large,Elsinga16POF,Carter17JFM,soulard2017influence}. The power-law scaling of wavelet spectra has also been used to gauge anisotropy \cite{horbury2008anisotropic,wicks2011anisotropy}. Our motivation here is to quantify anisotropy over all scales and in flows that may be inhomogeneous. 

Fourier analysis is fraught with complications when applied to inhomogeneous fields \cite{Sadek18}. Afterall, Fourier modes are not an eigenbasis for arbitrary domains and boundary conditions \cite{champeney1987handbook,krantz2019panorama}. Measuring the spectrum via a Fourier transform of the auto-correlation function, sometimes known as the Wiener-Khinchin theorem \citep{champeney1987handbook}, is not justified in the presence of boundaries or if the field is statistically inhomogeneous such as with a spatially varying mean or autocorrelation. 
In practical applications, Fourier analysis of inhomogeneous fields (or non-stationary temporal signals) is often performed \cite{oppenheim1999signalprocessing,ThomsonEmery01} after removing the ensemble-mean \cite{soulard2012inertial}, detrending \cite{Savage+2017,ORourke2018}, and/or tapering (\textit{i.e.} windowing) \cite{Scott2005,khatri2018surface}. Doing so removes potentially important components of the dynamics. An emblematic example is the global oceanic circulation, for which it had been asserted since the advent of global satellite altimetry in the 1990s that its wavenumber spectrum's peak is at scales $O(100)~$km based on detrended and windowed Fourier analysis (e.g. \cite{FerrariWunsch09,Torresetal2018jgr,Kleinetal2019}). It was recently shown \cite{Storer2022NatComm,buzzicotti2023spatio} that this is untrue and that the spectral peak is in fact at $O(10^4)~$km. The spectral peak and the existence of a power-law scaling over scales $>10^3~$km in the oceanic circulation could not have been detected from windowed Fourier analysis because all scales larger than the window size (typically taken to be a few hundred kilometers to avoid continental boundaries and curvature effects) are implicitly removed. These limitations of Fourier analysis exist for many realistic flows, including the Rayleigh-Taylor flows we consider here. 

When removing the mean flow before using Fourier analysis, a starting assumption is that identifying the statistical ensemble in an application is well-posed. However, the choice of an ensemble is seldom unique and can have important implications on the conclusions as discussed in the case of thermal convection by Kraichnan \cite{kraichnan1964direct}. For Earth systems applications, such in the case of oceanic or atmospheric observations, we only have a single realization. Appealing to ergodicity with respect to time to treat the temporal record as an ensemble is formally unjustified (although it is often employed) due to a lack of statistical stationarity: the system is variable over a wide range of time-scales. Even if we overlook the difficulties in choosing an ensemble, and if fluctuations about the ensemble-mean flow were assumed to be statistically homogeneous, scales of the ensemble-mean flow itself would remain inaccessible \cite{grea2013rapid}. Often, there is a lack of scale separation between the mean and fluctuating flow components. In such systems, the ensemble-mean flow has significant spatial (or temporal) variations at the same scales populated by the fluctuating field \cite{buzzicotti2023spatio}. This necessitates a self-consistent scale analysis of both the mean and fluctuating contributions to the full non-linear dynamics.

Another tool for analyzing scales is the 2nd-order structure function. It has been a valuable phenomenological tool in turbulence theory, but it requires statistical averaging and is not a formal scale decomposition of a field \cite{Frisch95}. Unlike a spectrum, which when integrated yields total energy (Parseval's relation), a 2nd-order structure function, $S_2(r)$, yields total energy in the limit\footnote{While the sum of the 2nd-order structure function and the autocorrelation yields total energy, the sum lacks scale information and is not a scale decomposition. If Fourier analysis is justified, such as for homogeneous flows, it is possible to relate the 2nd-order structure function, $S_2(r)$, to the Fourier spectrum, $E(k)$, but this follows directly from the Wiener-Khinchin relation and involves a weighted average of $E(k)$ over the \emph{entire} $k$-space, $S_2(r)=2\int_0^\infty dk\,(1-cos(k\,r))E(k)$ (e.g. \cite{babiano1985structure,Pope2001}).} of length-scales $r\to0$ (e.g. \cite{balwada2022direct}). At any scale $r$, $S_2(r)$ can have significant contributions from all scales larger or smaller than $r$. This is unsurprising since for a field such as velocity $u(x)$, $S_2(r)=\langle|\delta u(x;r)|^2\rangle$ at scale $r$ is constructed from increments $\delta u(x;r)=u(x+r) - u(x)$ of separation $r$ before spatial averaging, $\langle\dots\rangle$. Increments $\delta u(r)$ can have contributions from all scales larger or smaller than $r$ depending on the regularity (or smoothness) of the field $u(x)$ \cite{eyink2005locality,Eyink09,Aluie17} (see discussion following eq.~(4) in \cite{aluie2010scale}). It is known that the power-law scaling of a 2nd-order structure function, $S_2(r)\sim r^{\alpha}$, is related to that of the Fourier spectrum, $E(k)\sim k^{-\alpha-1}$, but only if $\alpha<2$, \textit{i.e.} the scaling relation breaks down if $E(k)$ is steeper than $k^{-3}$ as a function of wavenumber $k$ (e.g. \cite{babiano1985structure,biferale2001inverse}). Perhaps less well-known is that the scaling relation between $S_2(r)$ and $E(k)$ also breaks down when $\alpha < 0$, \textit{i.e.} $E(k)$ is shallower than $k^{-1}$ \cite{Eyink95,eyink2005locality,aluie2010scale}. Figs.~\ref{fig:1D_fields_with_scaling}-\ref{fig:1D_fields_structure_func} in the Appendix provide a demonstration of these facts.
Another obvious limitation, shared with Fourier analysis, is that structure functions do not provide spatial information about various scales.

Of relevance to results herein are KE spectra of variable density (VD) flows \cite{livescu2008variable,Livescu20}, characterized by significant variations in mass density $\rho$. When using Fourier analysis, structure functions, or wavelets to analyze KE, $\rho|\bu|^2/2$, as a function of scale, KE is often treated as quadratic to ensure the spectrum (\textit{i.e.} power spectral density) is positive. For example, the Fourier transform of an auxiliary field $\bw\equiv \sqrt{\rho}\bu$ is performed and then squared such that the spectrum is positive and integrates to $\rho|\bu|^2/2$ \cite{kida1990energy,CookZhou02}. This approach, while being mathematically sound on its own, creates difficulties when  analyzing the scale-dependent equations governing such KE spectra as shown in \cite{zhao2018inviscid}. 

Recently, the so-called `filtering spectrum' was proposed \cite{Sadek18} to measure spectral content using straightforward coarse-graining in physical space, which is closely related to the continuous wavelet transform \cite{daubechies1992ten,perrier1995wavelet}. This permits its application to inhomogeneous flows with complex boundaries and allows us to probe scales of both the mean and fluctuating fields concurrently \cite{buzzicotti2023spatio}.
The approach has been recently adopted to measure the first global energy spectrum of the oceanic general circulation \cite{Storer2022NatComm}. 

The filtering spectrum can be regarded as a generalization of the Fourier spectrum to inhomogeneous fields. The filtering spectrum
is an energy-preserving scale decomposition \cite{Sadek18} and can represent the non-quadratic KE content at different scales of VD flows as shown in \cite{Zhao22JFM}. If the filtering kernel has a sufficient number of vanishing moments, the filtering spectrum follows any power-law scaling that the Fourier spectrum may have (assuming Fourier analysis is possible). In fact, the filtering spectrum converges to the Fourier spectrum when using a kernel with an infinite number of vanishing moments (e.g. the Dirichlet kernel), which is justified only for homogeneous fields given the highly non-local nature of such kernels in $x$-space (Fig.~\ref{fig:FilterKernels} in Appendix). 

In the appropriate domain, Fourier basis functions satisfy orthogonality, which is a highly prized property. However, the sum of any finite number of these basis functions\footnote{An equivalent statement can be made when Fourier modes are continuous.} suffers from significant spatial non-locality\footnote{Non-locality of the Dirichlet kernel in $x$-space is a consequence of non-smoothness of its Fourier transform (the sharp-spectral cutoff) in $k$-space, which is a consequence of what is sometimes known as the Paley–Wiener theorem \cite{sogge2017fourier,krantz2019panorama}.}. Such spatial non-locality is not a concern when working with homogeneous fields, but presents significant challenges when working with inhomogeneous fields such as in the presence of boundaries (see Fig.~\ref{fig:FilterKernels} and associated discussion in Appendix). What has been shown in previous work \cite{Sadek18,Raietal2021,Zhao22JFM,Storer2022NatComm,buzzicotti2023spatio} is the possibility of performing a meaningful scale decomposition of inhomogeneous fields and determine their spectra, satisfying both positive semi-definiteness and energy conservation, without the need for orthogonality.

The filtering spectrum as a method is especially valuable in permitting us to visualize (in physical space) the flow at different scales in a self-consistent manner \cite{buzzicotti2021}, along with any associated anisotropy as we shall see below. A disadvantage of the filtering spectrum compared to the Fourier spectrum is that it involves smoothing as a function of scale \cite{Sadek18}. This is the price paid for gaining spatially local information at different scales and generalizing the notion of a spectrum to non-homogeneous fields. Concurrently exact spatial and scale localization is forbidden by the uncertainty principle \cite{sogge2017fourier,krantz2019panorama}.

The following section~\ref{sec:rt_spectra} is a brief review of the 1D filtering spectrum proposed by \cite{Sadek18}. In section~\ref{sec:MultiDspectra}, we generalize the filtering spectrum to multiple dimensions and define simple metrics to quantify scale anisotropy. In section~\ref{sec:numerics}, we demonstrate our approach using illustrative examples and then apply it to anisotropic inhomogeneous turbulence generated from the Rayleigh-Taylor (RT) instability in 2D and in 3D. The paper closes with a brief summary and practical comments about the approach's usage.

\section{One-dimensional filtering spectrum} \label{sec:rt_spectra}
For any field $\ba(\bx)$, a coarse-grained or (low-pass) filtered version of this field, which contains spatial variations
at scales $>\ell$, is defined in $n$-dimensional Euclidean space\footnote{Coarse-graining on curved manifolds is more involved \cite{aluie2019convolutions}.} as \cite{Leonard75,Germano92,MeneveauKatz00,eyink2005locality}
\be
\OL \ba_\ell(\bx) = \int \mathrm{d}^n\br~ G_\ell(\bx-\br)\, \ba(\br).
\lb{eq:filtering_1}\ee
Kernel $G_\ell(\br)= \ell^{-n} G(\br/\ell)$ is the dilated version of the ``parent kernel'' $G(\br)$, which is normalized.  $G_\ell(\br)$ has its main support over a region of diameter $\ell$. 
Operation (\ref{eq:filtering_1}) may be interpreted as a local space average in a region of size $\ell$ centered at point $\bx$. It is, therefore, a scale decomposition performed in x-space that partitions length scales in the system into large ($\gtrsim\ell$), captured by $\OL \ba_\ell$, and small ($\lesssim\ell$), captured by the residual 
\be
\ba'_\ell=\ba-\OL\ba_\ell.
\ee
We assume $G_\ell$ is an even function such that, $\int \br\, G_\ell(\br) \, \mathrm{d}^n \br = 0$, which ensures that local averaging is symmetric and operation~\eqref{eq:filtering_1} can be rewritten as
\be
\OL \ba_\ell(\bx) = \int \mathrm{d}^n\br~ G_\ell(\br)\, \ba(\bx+\br).
\lb{eq:filtering}\ee
Coarse-graining is a very general scale-analysis framework and includes Fourier analysis (e.g. \cite{Krantz99,Aluie09}) and wavelet analysis (e.g. \cite{meneveau1991analysis,meneveau1991dual}) as special cases with the appropriate choice of kernel $G_\ell$. See \cite{Sadek18} for further discussion. Coarse-graining also lays the foundational framework of large-eddy simulation (LES) \citep{Leonard75,Germano92,MeneveauKatz00,GhateLeLe20JFM,Johnson22JFM}.
For inhomogeneous flows such as in the presence of boundaries, it is traditional in LES to filter only along the homogeneous directions (e.g. \cite{Piomellietal91,yang2015integral}). This is to prevent commutation errors that arise when deforming the kernel to avoid the boundary \cite{SagautLES}. However, at least for the purpose of diagnosing scales, it was shown recently that these considerations are moot if regions beyond boundaries are treated in a manner that is consistent with the boundary conditions satisfied by the dynamics \cite{zhao2018inviscid,Aluieetal2018_jpo,rai2021scale,buzzicotti2023spatio}.
We will discuss this point in greater detail when applying the method to data from Rayleigh-Taylor simulations.

Since spatial coarse-graining  allows for extracting spatial information at varying length scales $\ell$, it can be used for calculating the one-dimensional filtering spectrum \cite{Sadek18}.
As an example, the filtering spectrum for the velocity field $\bu$
is
\begin{align} 
\OL{E}(k_\ell)= \frac{d}{dk_\ell}\left\{ \frac{|\OL{\bu}_\ell(\bx)|^2}{2}\right\},
\label{eq:FilteringSpectrum}
\end{align}
where $k_\ell=L/\ell$ is a `filtering wavenumber', $L$ is a characteristic length-scale (e.g. domain size),  $\ell$ is the scale being probed, and $\{\dots\} \equiv (\mbox{Volume})^{-1}\int (\dots) d^n\br$ is spatial averaging in $n$-dimensions. 
Eq.~\eqref{eq:FilteringSpectrum} measures the energy density (per wavenumber) at scale $\ell$ by varying it and probing the associated variations in coarse KE, $\{ |\OL{\bu}_\ell(\bx)|^2\}/2$, which is the cumulative spectrum at \emph{all} scales larger than $\ell$. The main advantage of this method is that it does not rely on Fourier transforms and, therefore, can be easily applied to non-periodic or homogeneous data. In a periodic domain,  Fourier and filtering spectra agree if $G_\ell$ has sufficient vanishing moments. In fact, the two spectra have an explicit relationship expressed by eq.~(16) in \cite{Sadek18}. Another advantage to the filtering spectrum is its ease of generalization to VD flows \cite{Sadek18,Zhao22JFM} where KE, $\rho|\bu|^2/2$, is non-quadratic due to variations in the mass density field, $\rho$. For VD flows, such as Rayleigh-Taylor flows considered in this paper, the KE filtering spectrum is \cite{Sadek18}
\begin{align} 
\OL{E}(k_\ell)\equiv \frac{d}{dk_\ell}\left\{\frac{1}{2}\frac{|\OL{\rho\bu}_\ell(\bx)|^2}{\OL\rho_\ell}\right\},
\label{eq:FilteringSpectrumVD}
\end{align} 
which reduces to eq.~\eqref{eq:FilteringSpectrum} when $\rho$ is constant.
Energy spectra based on Fourier and wavelet transforms, on the other hand, are constrained to quadratic quantities to ensure that the spectrum is positive while also satisfying Parseval's relation. Therefore, a Fourier or wavelet decomposition usually treats kinetic energy as quadratic by transforming the auxillary field $\sqrt{\rho}\bu$, even when this treatment obscures the inertial-range dynamics \citep{zhao2018inviscid}. In contrast, the filtering spectrum conserves energy due to the fundamental theorem of calculus and is not constrained by Parseval's relation. Figure 5 in reference \cite{Zhao22JFM} shows the density,
velocity, and kinetic energy filtering spectra of 2D and 3D RT turbulence obtained with this new approach. 




\section{Multi-dimensional filtering spectra and scale anisotropy}
\lb{sec:MultiDspectra}
While the filtering spectrum in eqs.~\eqref{eq:FilteringSpectrum},\eqref{eq:FilteringSpectrumVD} can be applied to anisotropic fields as was done in \cite{Zhao22JFM}, it is a one-dimensional spectrum since it does not distinguish between scales along different directions. The contribution from the most energetic scale direction dominates the value of $\OL{E}(k_\ell)$. In order to distinguish the spectral content in different directions, we  generalize the above definition to multiple dimensions by using anisotropic kernels. 

\begin{figure}
\centering
\begin{minipage}[b]{1.0\textwidth}
\centering
    \subfigure[Multi-D filtering spectrum in eq.~\eqref{eq:anisotropic_spectra}]
{\includegraphics[height=0.42\textwidth]{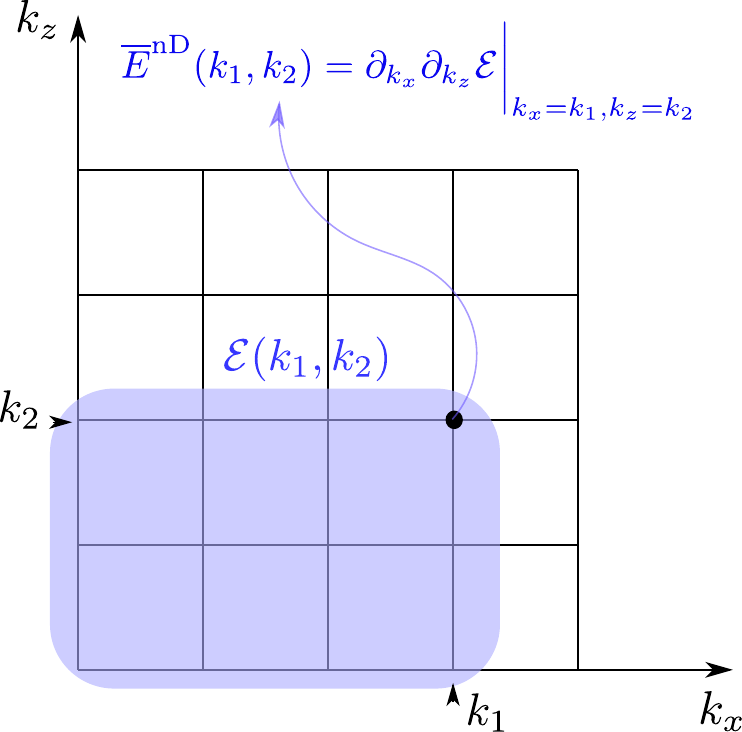}\label{fig:3DfilteringSpec}}
\phantom{}
    \subfigure[Anisotropy based on shell centroid]{\includegraphics[width=0.4\textwidth]{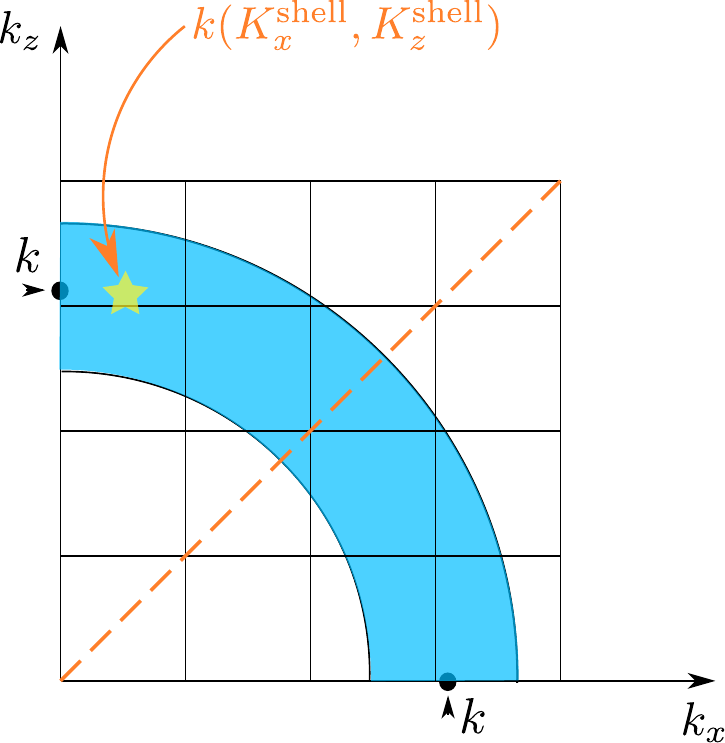}}
    \caption{{\textbf{Panel (a):}} Schematic of extracting the multi-dimensional ``filtering spectrum''. Filtering is performed in physical space to disentangle scales, without resorting to Fourier transforms.  Filtering wavenumbers $(k_x,k_z)=(L/\ell_x,L/\ell_z)$ are only a proxy for length-scales, but may be thought of as Fourier wavenumbers conceptually. The cumulative spectrum $\mathcal{E}(k_1, k_2)$ (shaded in blue) yields the energy content at \emph{all} scales larger than $\ell_1$ in the x-direction and larger than $\ell_2$ in the z-direction, such that $\partial_{k_z}\partial_{k_x}\mathcal{E}$ evaluated at $(k_1, k_2)$ yields the spectral energy (density) content at exactly those scales $(\ell_1,\ell_2)$. It is a generalization of the traditional multi-dimensional Fourier spectrum, $\frac{1}{2}|\wh{\bu}|^2(k_1,k_2)$, and reduces to it when using a sharp-spectral filtering kernel. 
{\textbf{Panel (b):}} Shell (blue) in filtering wavenumber space, $|\bk| \in [k/\sqrt{2}, \sqrt{2}k]$. It has logarithmic thickness $\times 2$ \citep{Aluie09}. Its centroid (yellow star) based on the distribution of $\OL{E}^{n\mathrm{D}}(\bk)$ within the shell ($\bK^\mathrm{shell}(k)$ in eq.~\eqref{eq:integral_K_ann}) gives a measure of anisotropy at scale $\ell = L/k$ from its proximity to the diagonal line (dashed orange). 
    \label{fig:AnisotropySpectraSchema_shell}
    }
\end{minipage}
\end{figure}

This is most easily done when considering multi-dimensional filtering kernels that are separable. For example in 3D with Cartesian coordinates,
\be
G_\ell(\br) = G_\ell(r_x)\, G_\ell(r_y) \, G_\ell(r_z), 
\lb{eq:separable}\ee
where $\br=(r_x,r_y,r_z)$, which is the case for a Gaussian or Boxcar kernel. For separable kernels, isotropic filtering in eq.~\eqref{eq:filtering} reduces to filtering in each direction separately:
\be
\OL{\bu}_\ell(\bx) = \int d\br\, G_\ell(\br)\ \bu(\bx+\br)= \int dr_x \,G_{\ell}(r_x) \int dr_y\, G_{\ell}(r_y) \int dr_z \,G_{\ell}(r_z)\ \bu(\bx+\br) ~.
\ee
Therefore, a natural generalization to anisotropic filtering is
\be
\OL{\bu}_{\vell}\,(\bx) = \int dr_x \,G_{\ell_x}(r_x) \int dr_y\, G_{\ell_y}(r_y) \int dr_z \,G_{\ell_z}(r_z)\ \bu(\bx+\br) ~.
\lb{eq:AnisoFiltering}\ee
This definition makes it clear how we can probe length scales along different directions by utilizing a vector of filtering scales, $\vell=(\ell_x,\ell_y,\ell_z)$. The ``cumulative spectrum'' 
\begin{align} \label{eq:cumu_filtered_energy}
\mathcal{E}(\bk) \equiv \frac{1}{2}\left\{ \frac{|\OL{\rho\bu}_{\vell}|^2}{\OL{\rho}_{\vell}}\right\}
\end{align}
at filtering wavevector 
\be
\bk=(k_x,k_y,k_z) \equiv (L/\ell_x,L/\ell_y,L/\ell_z)
\lb{eq:filteringwavevector}\ee 
yields the energy content at all
scales larger than $\ell_i$ in the $i$th-direction. Here, $L$ is a reference scale common to all directions. Fig.~\ref{fig:3DfilteringSpec} illustrates the cumulative spectrum in 2-dimensions, $\mathcal{E}(k_1, k_2)$, which is shaded in blue and accounts for all energy at filtering wavevectors $|k_x|<k_1$ and $|k_y|<k_2$. Note that $\bk$ in eq.~\eqref{eq:filteringwavevector} occupies only the first quadrant, $k_x,k_y,k_z\ge0$. This is because $\mathcal{E}(\bk)$ in eq.~\eqref{eq:cumu_filtered_energy} is invariant under rigid (Euclidean isometry) transformations of the fields $\rho(\bx)$ and $\bu(\bx)$, which involve a sequence of rotations, translations, and reflections. Phase information, which may seem lost in the definition of $\mathcal{E}(\bk)$, is easily retrieved from the spatial field $\frac{1}{2}|\OL{\rho\bu}_{\vell}|^2(\bx)/\OL{\rho}_{\vell}(\bx)$ in eq.~\eqref{eq:cumu_filtered_energy} without spatial averaging, $\{\dots\}$.

From $\mathcal{E}(\bk)$, we can quantify the spectral energy density at any scale $\vell$ in $n$-dimensions by defining the $n$-dimensional filtering spectrum as
\begin{align} \label{eq:anisotropic_spectra}
    \OL{E}^{n\mathrm{D}}(\bk) \equiv \frac{\partial^n}{\partial k_1\dots \partial k_n}  \mathcal{E}(\bk)~.
\end{align}
It satisfies energy conservation in a straightforward manner by the fundamental theorem of calculus, 
\be
\frac{1}{2}\left\{ \frac{|\OL{\rho\bu}_{\vell_0}|^2}{\OL{\rho}_{\vell_0}}\right\} + \int_{k_0}^\infty \md^n \bk  ~\OL{E}^{n\mathrm{D}}(\bk) = \frac{1}{2}\left\{ \rho |\bu|^2\right\} ~,
\lb{eq:EnergyConserve}\ee
where $\vell_0 = (L/k_0,L/k_0,L/k_0)$ is the largest length scale at which the flow is filtered. 

Fig.~\ref{fig:3DfilteringSpec} illustrates the motivation for our definition~\eqref{eq:anisotropic_spectra} in 2D. The filtering spectrum
$\OL{E}^{2\mathrm{D}}(k_1, k_2)$ at filtering wavevector $\bk=(k_1,k_2)$ quantifies the spectral energy density by measuring the cumulative spectrum's response, $\partial_{k_z}\partial_{k_x}\mathcal{E}(k_1,k_2)$, to concurrent scale variations in all directions, $\Delta k_x$ and $\Delta k_z$, at $(k_1, k_2)$. Here, the filtering scale vector is $(\ell_1,\ell_2) = (L/ k_1, L/ k_2)$. The one-dimensional spectrum $\OL{E}(k_\ell)$ in eq.~\eqref{eq:FilteringSpectrumVD} is essentially an integral of $\OL{E}^{2\mathrm{D}}(\bk)$ at all filtering wavevectors $\bk=(k_x,k_y)$ within a thin shell of width $\Delta k$, $\left|\sqrt{k_x^2+k_y^2}-k_\ell\right|<\Delta k$.

When mass density $\rho$ is constant, $\OL{E}^{n\mathrm{D}}(\bk)$ is a generalization of the traditional  Fourier spectrum,
$\frac{1}{2}|\wh{\bu}|^2(\bk)$, where $\wh{\bu}(\bk)$ is the Fourier transform of $\bu(\bx)$. The filtering spectrum reduces to the Fourier spectrum when using a sharp-spectral filtering (Dirichlet or Bessel) kernel in eq.~\eqref{eq:AnisoFiltering}. However, a main advantage 
of our multi-dimensional filtering spectrum in equation~(\ref{eq:anisotropic_spectra}) is that it can diagnose bounded or inhomogeneous flows in a straightforward manner.
Moreover, as discussed in section~\ref{sec:rt_spectra},
it allows for measuring the spectrum of KE in VD flows while 
respecting its nonquadratic nature.

\subsection{Scale-dependent Metrics for Shape Anisotropy}
The multi-D filtering spectrum allows us to quantify anisotropy as a function of length-scale as follows. For any filtering wavenumber $k$ associated with length scale $\ell$, consider the logarithmic shell $|\bk| \in [k/\sqrt{2}, \sqrt{2}k]$ in filtering wavevector space shown in Fig.~\ref{fig:AnisotropySpectraSchema_shell}b. Within the shell, the normalized ``scale-dependent centroid'' (or first moment) of $\OL{E}^{n\mathrm{D}}$ is
\begin{align} \label{eq:integral_K_ann}
\bK^\mathrm{shell}(k)\equiv
\begin{cases}
         \left.\displaystyle\frac{1}{k}\displaystyle\int\limits_{|\bk| \in \left[\sfrac{k}{\sqrt{2}}, \sqrt{2}k\right]}\md^n\bk\, \bk \,\OL{E}^{n\mathrm{D}}(\bk)~\middle/\displaystyle\int\limits_{|\bk| \in \left[\sfrac{k}{\sqrt{2}}, \sqrt{2}k\right]}\md^n\bk \, \OL{E}^{n\mathrm{D}}(\bk)~\right.,\\\\
         \hspace{.5cm}\bzed \hspace{1cm}\text{if $\displaystyle\int\limits_{|\bk| \in \left[\sfrac{k}{\sqrt{2}}, \sqrt{2}k\right]}\md^n\bk \, \OL{E}^{n\mathrm{D}}(\bk) < \epsilon^{\textrm{numeric}}$.}
\end{cases}
\end{align}
Integration is performed in the first quadrant of $\bk$-space. Definition~\eqref{eq:integral_K_ann} is dimensionless and is zero if the shell is devoid of energy. We set $\epsilon^{\textrm{numeric}}=10^{-14}$ in this work to avoid numerical overflow errors. When multiplied by wavenumber, $k\,\bK^\mathrm{shell}(k)$ is the centroid for the shell's energy as shown in Fig.~\ref{fig:AnisotropySpectraSchema_shell}b.
Note the shells $\left[\sfrac{k}{\sqrt{2}}, \sqrt{2}k\right]$ in definition~\eqref{eq:integral_K_ann} have a width that is constant on a logarithmic (not linear) scale. This is important to capture spatially localized structures at different scales, sometimes called a Littlewood-Payley decomposition \cite{Aluie09}.

\begin{figure}
\centering
\begin{minipage}[b]{1.0\textwidth}
\centering
    \subfigure{\includegraphics[width=0.5\textwidth]{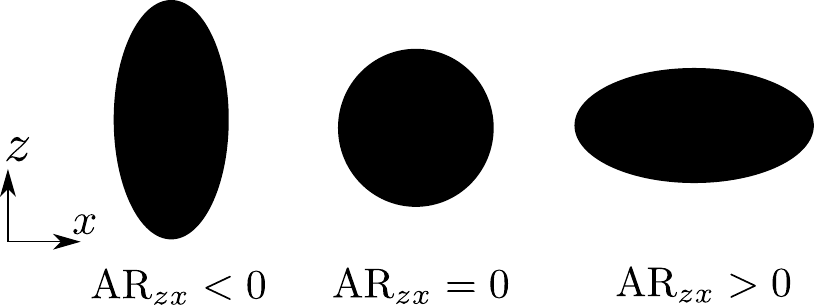}}
    \caption{Anisotropic shapes and their corresponding anisotropy metric $\mathrm{AR}$ from eqs.~(\ref{eq:AR_ann}),(\ref{eq:AR_mar}).
\label{fig:Eddies_AR_diagram}}
\end{minipage}
\end{figure}

The centroid's distance from the diagonal in Fig.~\ref{fig:AnisotropySpectraSchema_shell}b is one measure of anisotropy at each $k$, which is equivalent to the contrast between different components, ${K}_i^\mathrm{shell}$, of $\bK^\mathrm{shell}$. A convenient measure for such contrast is what we call the scale-dependent anisotropy metric (AR) tensor,
\begin{align} \label{eq:AR_ann}
\mathrm{AR}_{ij}^\mathrm{shell}(k) \equiv \frac{({K}_i^\mathrm{shell})^2-({K}_j^\mathrm{shell})^2}{({K}_i^\mathrm{shell})^2+({K}_j^\mathrm{shell})^2}~, \hspace{1cm}\mathrm{for~}i,j=1,\dots,n.
\end{align}
If ${K}_i^\mathrm{shell}={K}_j^\mathrm{shell}=0$, we define $\mathrm{AR}_{ij}^\mathrm{shell}\equiv0$. For example, $\mathrm{AR}_{zx}^\mathrm{shell}(k)>0$, indicates a dominant concentration of energy within shell $k$ at small $k_x$ and large $k_z$. Therefore, $\mathrm{AR}_{zx}^\mathrm{shell}(k)>0$ indicates structures that are horizontally elongated (along the $x$-direction) at scale $\ell \sim k^{-1}$ as sketched in Fig.~\ref{fig:Eddies_AR_diagram}. The converse is also true, where $\mathrm{AR}_{zx}^\mathrm{shell}(k)<0$ indicates vertically elongated structures at scale $\ell \sim k^{-1}$. The anisotropy metric tensor is antisymmeteric, $\mathrm{AR}_{ij}^\mathrm{shell}=-\mathrm{AR}_{ji}^\mathrm{shell}$, and each of its components lies within $-1\le\mathrm{AR}_{ij}^\mathrm{shell}\le1$. We shall demonstrate these definitions numerically using simple examples in the following section.
In Appendix section~\ref{sec:MarginalSpectra}, we present a reduced (and less expensive) scale analysis based on what we call marginal spectra and discuss their limitations.

\section{Numerical Implementation}\lb{sec:numerics}
In this section, we measure scale anisotropy using illustrative examples as well as data from Rayleigh-Taylor (RT) turbulence simulations in 3D and 2D.
RT turbulence is markedly anisotropic and inhomogeneous, which makes it a good application for our method.


\subsection{Illustrative Examples}

\begin{figure}[bhp]
\centering 
\begin{minipage}[b]{1.0\textwidth}  
\centering
\subfigure
    {\includegraphics[width=0.35\textwidth]{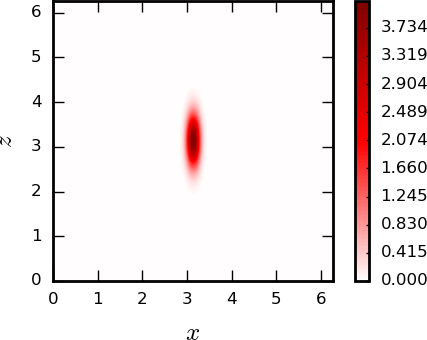} } \phantom{separation}
\subfigure
{\includegraphics[width=0.35\textwidth]{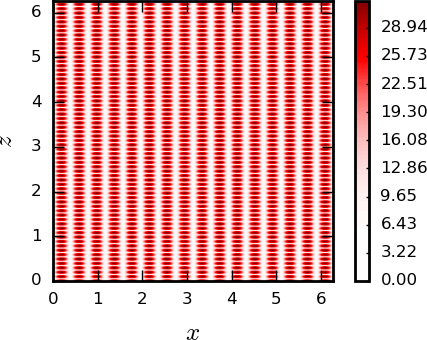}}
    \caption{Illustrative examples of anisotropic fields. Left panel shows the Gaussian scalar field in eq.~\eqref{eq:Gaussian}. Right panel shows
 the magnitude of the Taylor-Green velocity in eq.~\eqref{eq:TG}, where vortices are elongated along the $x$-direction.
 \label{fig:synthetic_data_viz}}
\end{minipage}
\end{figure}

For illustrative purposes, we consider two examples in a 2D periodic domain: a Gaussian scalar field and a Taylor-Green velocity field. Defining these in a periodic domain allows us to use a sharp spectral filtering kernel in Fourier space for comparison. The Gaussian scalar field $\phi(x,z)$ is of the form
\begin{align}
\lb{eq:Gaussian}
    \begin{split}
\phi(x,z) = &\frac{1}{\sqrt{2\pi}\sigma_x}\frac{1}{\sqrt{2\pi}\sigma_z}e^{-\frac{(x-\pi)^2}{2\sigma_x^2}-\frac{(z-\pi)^2}{2\sigma_z^2}}~,\\
        \mathrm{where}\hspace{1cm} &(\sigma_x,\sigma_z)=(\pi/32,\pi/8) \hspace{1cm}\mathrm{and}\hspace{1cm}x, z\in [0, 2\pi]~.
    \end{split}
\end{align}
This Gaussian field is anisotropic, elongated in the $z$-direction as shown in figure \ref{fig:synthetic_data_viz}(a). It populates a broad band of Fourier wavenumbers (see Appendix Fig.~\ref{fig:gaussian_fourier_spec}).

The Taylor-Green (TG) velocity field 
\begin{align}
\lb{eq:TG}
    \begin{split}
        (u_x,u_z) = &\left(p_z\sin\left(p_x x\right)\cos\left(p_z z\right),-p_x\cos\left(p_x x\right)\sin\left(p_z z\right)\right),\\
        \mathrm{where}\hspace{1cm} &(p_x,p_z)=(8,32) \hspace{1cm}\mathrm{and}\hspace{1cm}x, z\in [0, 2\pi]~,
    \end{split}
\end{align}
is anisotropic and elongated in the $x$-direction as shown in figure~\ref{fig:synthetic_data_viz}(b). It consists of only a single Fourier wavevector $(p_x,p_z)$. Both examples are discretized on a $512\times 512$ grid.


\begin{figure}[bhp]
\centering 
\begin{minipage}[b]{1.0\textwidth}  
\centering
\subfigure[~cumulative spectrum $\mE(k_x,k_z)$]
{\includegraphics[width=0.48\textwidth]{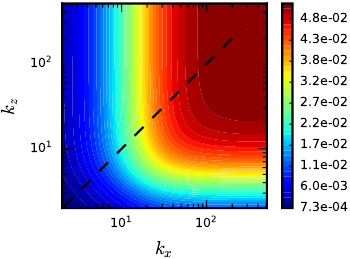} } 
\subfigure[~2D filtering spectrum $\OL{E}^{2\textrm{D}}(k_x,k_z)$]
{\includegraphics[width=0.48\textwidth]{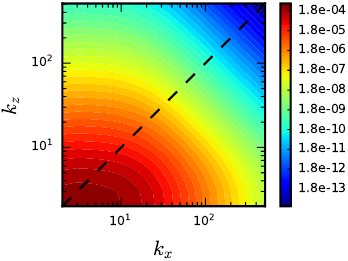}} \\
\subfigure[~scale-dependent centroid]
{\includegraphics[width=0.42\textwidth]{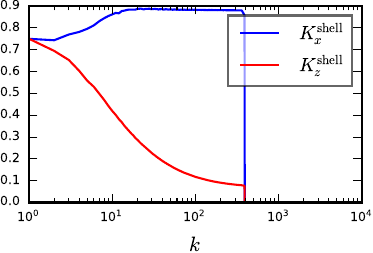} }
\subfigure[~scale-dependent anisotropy metric]
{\includegraphics[width=0.42\textwidth]{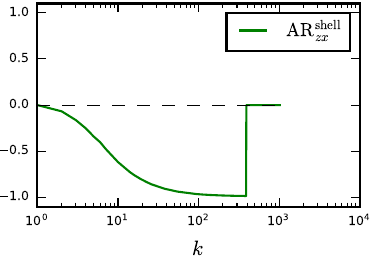} }
    \caption{Filtering spectra of the Gaussian scalar field (eq.~\eqref{eq:Gaussian}) using Gaussian filters (eq.~\eqref{eq:GaussKernel}). Panel~(a) is the cumulative spectrum in 2D filtering wavenumber space. Panel~(b) is the 2D filtering spectrum obtained  via eq.~\eqref{eq:anisotropic_spectra} (note the logarithmic color bar). Panel~(c) shows the scale-dependent centroid obtained from integrating $\OL{E}^{2\textrm{D}}$ within a shell (eq.~\eqref{eq:integral_K_ann}) . Panel~(d) shows the anisotropy metric from eqs.~\eqref{eq:AR_ann}.}
 \label{fig:anisotropic_gaussian}
\end{minipage}
\end{figure}

\begin{figure}[bhp]
\centering 
\begin{minipage}[b]{1.0\textwidth}  
\centering
\subfigure[$\OL{\phi}_{L/2}$]
{\includegraphics[width=0.42\textwidth]{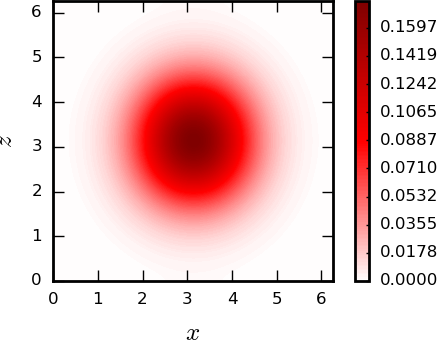} } 
\subfigure[$\OL{\phi}_{L/8}$]
{\includegraphics[width=0.42\textwidth]{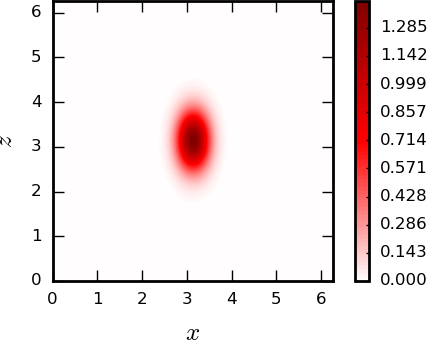}} \\
\subfigure[$\phi'_{L/8}$]
{\includegraphics[width=0.42\textwidth]{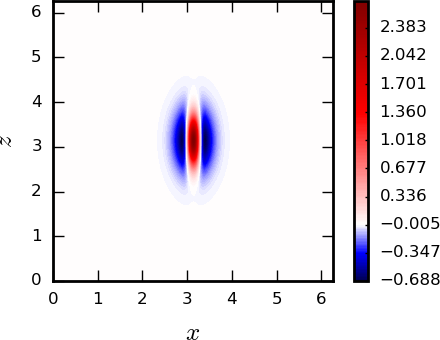} }
\subfigure[$\phi'_{L/32}$]
{\includegraphics[width=0.42\textwidth]{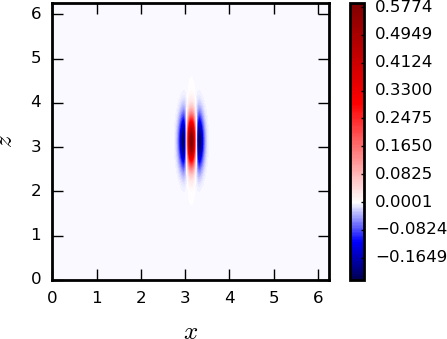} }
    \caption{Scale-dependent anisotropy demonstrated through the low-pass ($\OL\phi_\ell$) and high-pass ($\phi'_\ell$) filtered field in eq.~\eqref{eq:Gaussian} (see Fig.~\ref{fig:synthetic_data_viz}a). We use an isotropic Gaussian filtering kernel defined in eq.~\eqref{eq:GaussKernel}, with $\ell_x=\ell_z=\ell$. Note that $\phi$ is isotropic at very large scales and becomes increasingly anisotropic at smaller scales, with the smallest scales in panel (d) exhibiting the most anisotropy among those shown here. These panels  highlight the advantages of coarse-graining over Fourier space analysis or structure functions.
 \label{fig:Gaussian_filtered_Gaussian_viz}}
\end{minipage}
\end{figure}

\subsubsection{Gaussian scalar field}
Figure \ref{fig:anisotropic_gaussian} demonstrates the filtering spectrum, $\OL{E}^{2D}(k_x,k_z)$, and how to quantify anisotropy for the Gaussian scalar field, $\phi$. The cumulative spectrum, $\mathcal{E}(k_x,k_z)$, is calculated from eq.~\eqref{eq:cumu_filtered_energy} by setting $\rho=1$ and replacing $\bu$ with $\phi$. In defining filtering wavenumber, $k=L/\ell$, we use the domain size $L=2\pi$.
The filtering kernel in eq.~\eqref{eq:AnisoFiltering} is itself a normalized Gaussian of the form 
\be
G_\vell(r_x,r_z)=\left(\frac{6}{\pi\ell_x^2}\right)^{1/2} e^{-6|r_x/\ell_x|^2}  \left(\frac{6}{\pi\ell_z^2}\right)^{1/2} e^{-6|r_z/\ell_z|^2}~.
\lb{eq:GaussKernel}\ee
Fig.~\ref{fig:anisotropic_gaussian}(a) shows the cumulative spectrum, $\mathcal{E}(k_x,k_z)$, as a function of filtering wavevector $(k_x,k_z) = (L/\ell_x, L/\ell_z)$. 
It is clear from Fig.~\ref{fig:anisotropic_gaussian}(a) that $\mathcal{E}(k_x,k_z)$ is asymmetric with respect to the diagonal line, $k_x=k_z$.  $\mathcal{E}(k_x,k_z)$ 
saturates more quickly with increasing $k_z$ than with increasing $k_x$. Equivalent, albeit more intuitive information is conveyed by the filtering spectrum, $\OL{E}^{2D}(k_x,k_z)$, in Fig.~\ref{fig:anisotropic_gaussian}(b). It shows that $\OL{E}^{2D}(k_x,k_z)$ is asymmetric with respect to the diagonal, $k_x=k_z$, with more weight present at smaller $k_z$ (\textit{i.e.} larger vertical length scales $\ell_z\sim k_z^{-1}$). This is consistent with the elongated profile of $\phi$ along the $z$-direction relative to the $x$-direction. Note that the asymmetry in Fig.~\ref{fig:anisotropic_gaussian}(b) is only over an intermediate range of scales.
Symmetry seems to be recovered in the lower-left and upper-right corners of panel~\ref{fig:anisotropic_gaussian}(b), which we shall now discuss. 

Symmetry in the lower-left corner of Fig.~\ref{fig:anisotropic_gaussian}(b) is consistent with isotropy at the largest scales, comparable to the domain size, where filtering in eq.~\eqref{eq:AnisoFiltering} approaches a domain average, $\OL\phi_\vell \to \{\phi\}$ as $\vell \to (\infty,\infty)$. The spatial mean, $\{\phi\}$, is uniform and, thus, trivially isotropic. Fig.~\ref{fig:Gaussian_filtered_Gaussian_viz} demonstrates this in physical space, which also highlights the advantages of coarse-graining. It may be helpful to think of a length scale decomposition as an operation analogous to removing one's eyeglasses, blurring an image isotropically. If the blurring is only slight (small $\ell$ in filtering kernel $G_\ell$) as in Fig.~\ref{fig:Gaussian_filtered_Gaussian_viz}b, one can still detect any anisotropy present at the large scales. However, such anisotropy is undetectable if the blurring is severe ($\ell\to \infty$) as in Fig.~\ref{fig:Gaussian_filtered_Gaussian_viz}a. Fig.~\ref{fig:Gaussian_filtered_Gaussian_viz}c,d show how anisotropy of the field $\phi$ in eq.~\eqref{fig:Gaussian_filtered_Gaussian_viz} persists at smaller scales, which is explained by simple analysis in Appendix section\ref{sec:AppGaussianFilteringSpectrum}. How can this be reconciled with the seeming symmetry in the upper-right corner of Fig.~\ref{fig:anisotropic_gaussian}(b)?

Symmetry in the upper-right corner of Fig.~\ref{fig:anisotropic_gaussian}(b) is merely due to the finite grid resolution, which precludes fully resolving shells beyond a certain wavenumber. Indeed, the closer we approach the upper-right corner along the diagonal $k_x=k_z$, the less of the shell is resolved on either side of the diagonal. This is related to having isotropic numerical grid cells at the smallest scales. The analytical (non-discretized) $\phi$ exhibits persistent anisotropy to arbitrarily small scales as explained in Appendix section\ref{sec:AppGaussianFilteringSpectrum}. 

Fig.~\ref{fig:anisotropic_gaussian}c evaluates centroid components $K^{\textrm{shell}}_x(k)$ and $K^{\textrm{shell}}_z(k)$ from eq.~\eqref{eq:integral_K_ann}. It shows that $K^{\textrm{shell}}_z \le K^{\textrm{shell}}_x$ over the entire range of $k$, which implies that the field is elongated along the $z$-direction at all scales. This is also expressed by the anisotropy metric in Fig.~\ref{fig:anisotropic_gaussian}d, where $AR^{\textrm{shell}}_{zx}\le 0$ at all scales. As we discussed above, at the largest length scales ($k\to 0$), we see that $K^{\textrm{shell}}_z \approx K^{\textrm{shell}}_x$ and $AR^{\textrm{shell}}_{zx}\approx 0$, indicating isotropy at very large scales. A similar behavior occurs as $k\to \infty$, but this is due to numerical resolution limits.



\begin{figure}[bhp]
\centering 
\begin{minipage}[b]{1.0\textwidth}  
\centering
\subfigure[~cumulative spectrum $\mE(k_x,k_z)$]
{\includegraphics[width=0.48\textwidth]{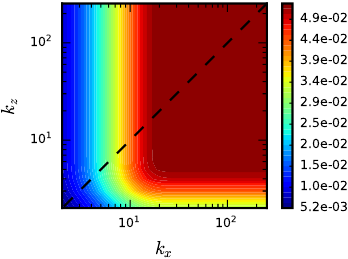} } 
\subfigure[~2D filtering spectrum $\OL{E}^{2\textrm{D}}(k_x,k_z)$]
{\includegraphics[width=0.48\textwidth]{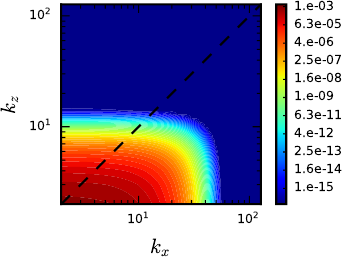}} \\
\subfigure[~scale-dependent centroid]
{\includegraphics[width=0.42\textwidth]{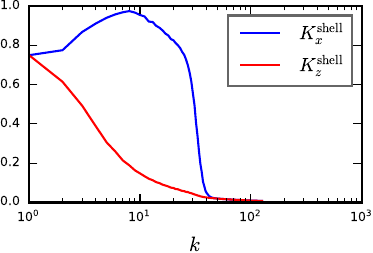} }
\subfigure[~scale-dependent anisotropy metric]
{\includegraphics[width=0.42\textwidth]{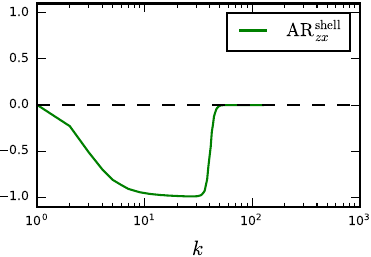} }
    \caption{Same as in Fig.~\ref{fig:anisotropic_gaussian} but filtering is done with a sharp-spectral kernel in Fourier space.
 \label{fig:anisotropic_gaussian_fourier}}
\end{minipage}
\end{figure}

Fig.~\ref{fig:anisotropic_gaussian_fourier} displays similar results as in Fig.~\ref{fig:anisotropic_gaussian} but using a sharp spectral filter. This is defined in Fourier space as
\begin{eqnarray}
\lb{eq:SharpSpectralFilter}
\wh{\mbox{Sinc}}_{\vec{k_c}}(k_x,k_z)=
\begin{cases}
1, &\text{if $|k_x|<k_{c,x}$ and $|k_z|<k_{c,z}$},\\
 0,  &\text{otherwise},
\end{cases}
\end{eqnarray}
which is straightforward to use due the domain's periodicity in our example. In eq.~\eqref{eq:SharpSpectralFilter}, $\vec{k_c}$ is the wavevector filtering cutoff.  $\wh{\mbox{Sinc}}_{\vec{k_c}}$ can be written in $x$-space as a product of Dirichlet kernels \cite{Riveraetal14} (see Appendix Fig.~\ref{fig:FilterKernels}).

Comparing Figs.~\ref{fig:anisotropic_gaussian},\ref{fig:anisotropic_gaussian_fourier} shows that the results are broadly consistent. Some notable differences are that plots in Fig.~\ref{fig:anisotropic_gaussian_fourier} are (i) sharper as a function of $k$ and (ii) decay more rapidly at large $k$. 

Differences in (i) sharpness arise from the filtering spectrum $\OL{E}(k_\ell)$ being essentially a weighted average of the Fourier spectrum $E(k)$ over a range of Fourier wavenumbers $k$ centered around $k_\ell$ \citep{Sadek18},
\be
\OL{E}(k_\ell) = \int_0^{\infty} \md p~ \left[\frac{d}{dk_\ell}  \left|\wh{G}\left(\frac{p}{k_\ell}\right)\right|^2\right] E(p).
\lb{eq:FilteringFourierSpectrum_Rel}\ee
Here, $\wh{G}$ is the Fourier transform of filtering kernel $G$. For a sharp spectral filtering kernel, the factor in square brackets in eq.~\eqref{eq:FilteringFourierSpectrum_Rel} is a delta function, yielding $\OL{E} = E$. Therefore, a comparison between Fig.~\ref{fig:anisotropic_gaussian_fourier} and Fig.~\ref{fig:anisotropic_gaussian} highlights that spatially localized filtering kernels such as the Gaussian are not strictly local in $k$-space compared to a sharp-spectral filter, which can lead to additional smoothing as a function of scale. This is a necessary price for gaining spatially local information (e.g. Fig.~\ref{fig:Gaussian_filtered_Gaussian_viz}, showing the field in $x$-space at different scales), since concurrently exact spatial and scale localization is forbidden by the uncertainty principle. 

The difference in (ii) the decay rate at large $k$ between Fig.~\ref{fig:anisotropic_gaussian_fourier} and Fig.~\ref{fig:anisotropic_gaussian}  was discussed in detail in \cite{Sadek18} and is a consequence of eq.~\eqref{eq:FilteringFourierSpectrum_Rel}. It is due to using the first-order kernel in eq.~\eqref{eq:GaussKernel} (using the terminology of \cite{Sadek18}) to calculate the filtering spectrum, which precludes $\OL{E}(k_\ell)$ decaying faster than $k_\ell^{-3}$ as $k_\ell\to\infty$. In contrast, the Fourier spectrum decays rapidly, $E(k)\sim e^{-k^2\sigma^2}$, for the Gaussian field $\phi$ in eq.~\eqref{eq:Gaussian} (see Appendix section\ref{sec:AppGaussianFilteringSpectrum}). The slower $k_\ell^{-3}$ decay is not a limitation of the filtering spectrum method, but of the filtering kernel itself. Faster decay rates can be captured by using higher-order kernels \cite{Sadek18} (Appendix Fig.~\ref{fig:1st_3rd_kernel}). Further discussion of this issue along with a comparison between Fourier and filtering spectra are provided in Appendix section\ref{sec:AppGaussianFilteringSpectrum}. 
Despite differences in the decay rate, we see that anisotropy inferred from spatial filtering in Fig.~\ref{fig:anisotropic_gaussian} is consistent with that in Fig.~\ref{fig:anisotropic_gaussian_fourier} from Fourier analysis. Therefore, even when using a first-order kernel, the underlying anisotropy is still detected from the filtering spectrum $\OL{E}^{n\textrm{D}}(k_x,k_z)$, although it may be underestimated. 


\subsubsection{Taylor-Green}
\begin{figure}[bhp]
\centering 
\begin{minipage}[b]{1.0\textwidth}  
\centering
\subfigure[~cumulative spectrum $\mE(k_x,k_z)$]
{\includegraphics[width=0.48\textwidth]{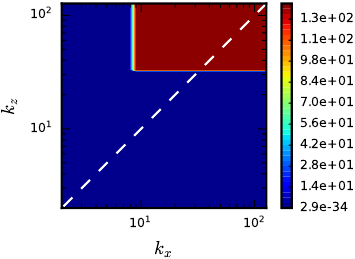} } 
\subfigure[~2D filtering spectrum $\OL{E}^{2\textrm{D}}(k_x,k_z)$]
{\includegraphics[width=0.48\textwidth]{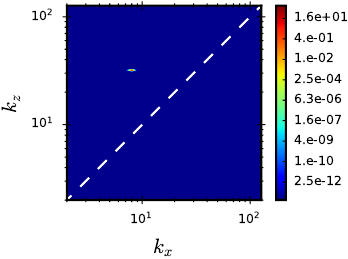}} \\
\subfigure[~scale-dependent centroid]
{\includegraphics[width=0.42\textwidth]{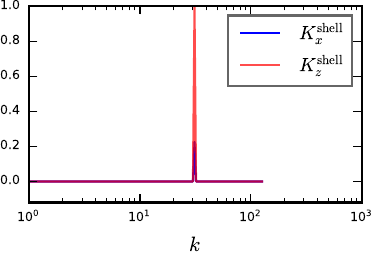} }
\subfigure[~scale-dependent anisotropy metric]
{\includegraphics[width=0.42\textwidth]{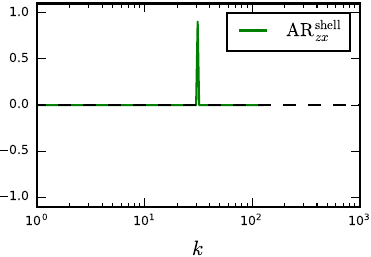} }
    \caption{Similar to Fig.~\ref{fig:anisotropic_gaussian} but analyzing the Taylor-Green velocity (eq.~\eqref{eq:TG}) using a sharp-spectral filter in Fourier space. The only Fourier mode present is $(k_x,k_z)=(8, 32)$, seen in panel (b).
    Panels (c) and (d) show the centroid (eqs.~\eqref{eq:integral_K_ann},\eqref{eq:integral_K_mar}), but with the integrals in eqs.~\eqref{eq:integral_K_ann},\eqref{eq:integral_K_mar} done over thin bands $k<|\bk|\leq k+1$ to highlight the precise wavenumber locations. A complementary figure~\ref{fig:anisotropic_TG_log_band} in the appendix shows integrals over dyadic bands $(k/\sqrt{2}, \sqrt{2}k]$ as in eqs.~\eqref{eq:integral_K_ann},\eqref{eq:integral_K_mar}. 
 \label{fig:anisotropic_TG}}
\end{minipage}
\end{figure}

Figure~\ref{fig:anisotropic_TG} shows the filtering spectrum, $\OL{E}^{2D}(k_x,k_z)$, and anisotropy for the Taylor-Green flow (eq.~\eqref{eq:TG}). For clarity, we use the sharp spectral kernel (eq.~\eqref{eq:SharpSpectralFilter}) for filtering since the flow is  localized at a single Fourier wavevector. We provide a corresponding Fig.~\ref{fig:anisotropic_TG_Gaussian} in the Appendix using the Gaussian filter for completeness.

In Fig.~\ref{fig:anisotropic_TG}a, the cumulative spectrum, $\mathcal{E}(k_x,k_z)$ from eq.~\eqref{eq:cumu_filtered_energy}, is calculated by setting $\rho=1$. In defining filtering wavenumber, $k=L/\ell$, we use the domain size $L=2\pi$.
It is clear from Fig.~\ref{fig:anisotropic_TG}(a),(b) that $\mathcal{E}(k_x,k_z)$ and $\OL{E}^{2\textrm{D}}(k_x,k_z)$ are asymmetric with respect to the diagonal line, $k_x=k_z$, with more weight toward smaller $k_x$ and larger $k_z$. In fact, $\OL{E}^{2\textrm{D}}(k_x,k_z)$ is zero everywhere except at $(k_x,k_z)=(8,32)$ as defined in eq.~\eqref{eq:TG}.

Fig.~\ref{fig:anisotropic_TG}(c) evaluates the normalized centroid $\bK^{\textrm{shell}}$ from eq.\eqref{eq:integral_K_ann}. Note that unlike in eq.\eqref{eq:integral_K_ann}, here we use thin bands $|\bk|\in(k,k+1]$ for clarity, since our field comprises of nonlocalized plane waves. For completeness, Fig.~\ref{fig:anisotropic_TG_log_band} in the Appendix uses the dyadic bands as in eq.\eqref{eq:integral_K_ann}.
Fig.~\ref{fig:anisotropic_TG}(c) shows a spike in $\bK^{\textrm{shell}}$ at $k=31$ with $K^{\textrm{shell}}_x = 0.23K^{\textrm{shell}}_z$ as expected, where the field is elongated along the $x$-direction. This is also expressed by the anisotropy metric in Fig.~\ref{fig:anisotropic_TG}d, where $\textrm{AR}^{\textrm{shell}}_{zx}(31)=0.90$.

\clearpage
\subsection{Rayleigh-Taylor Turbulence}

\begin{figure}[bhp]
\centering 
\begin{minipage}[b]{1.0\textwidth}  
\centering
\subfigure
    {\includegraphics[width=0.2\textwidth]{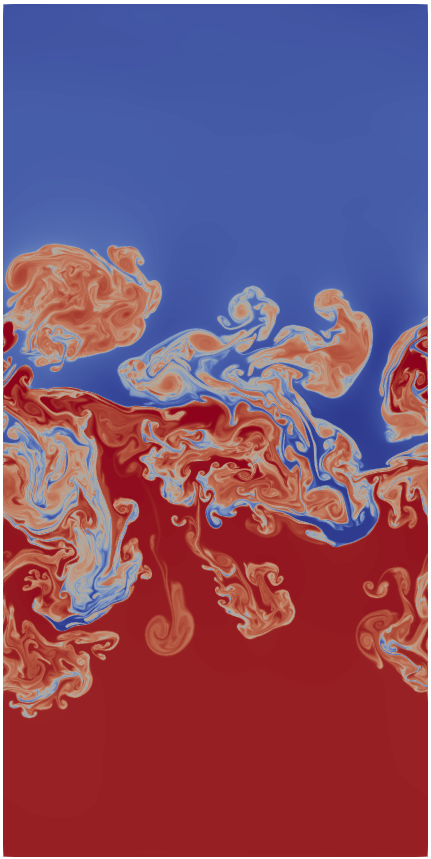} } \phantom{separate}
    {\includegraphics[width=0.4\textwidth]{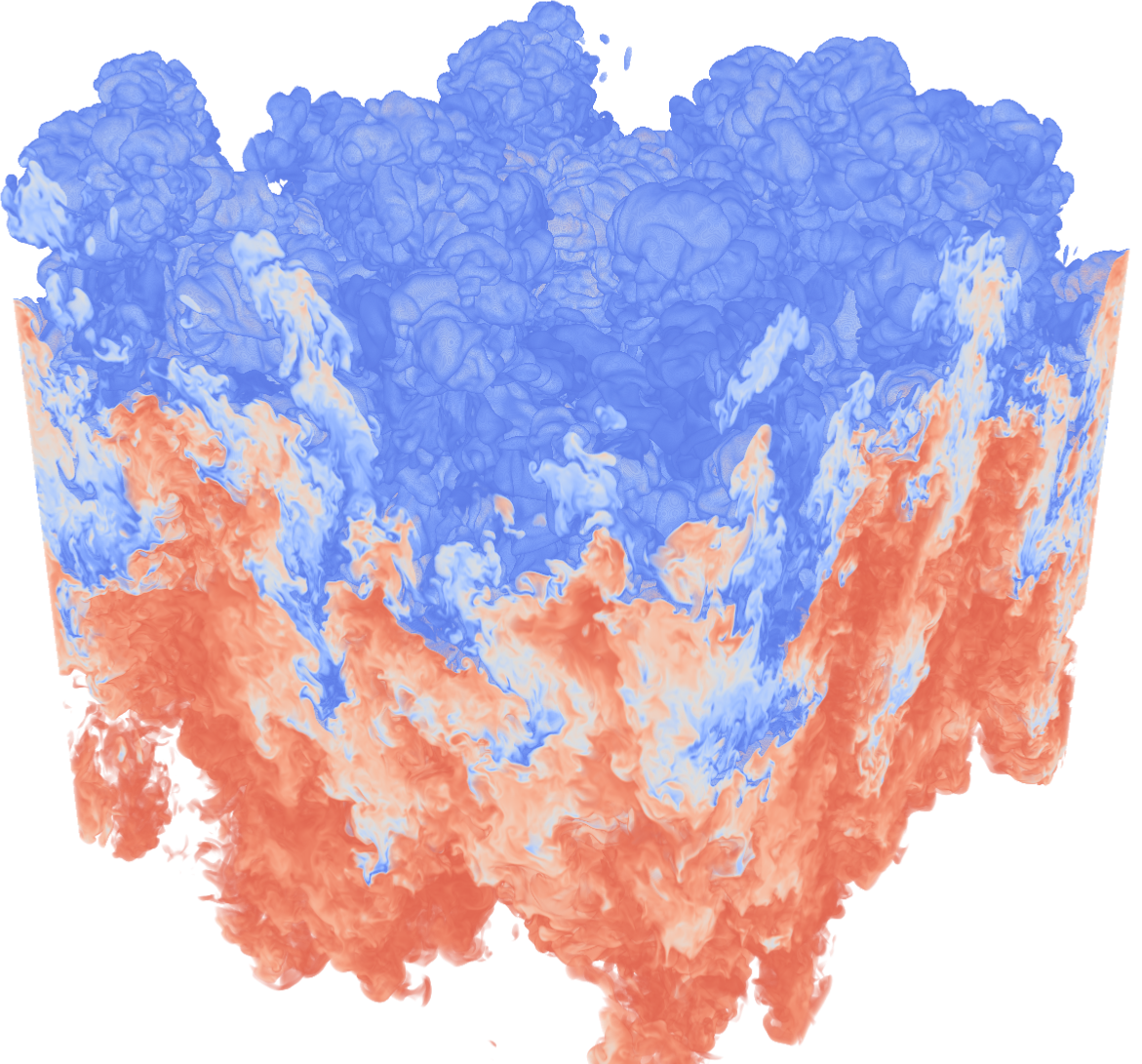} } 
    \caption{Visualizations of the density field from 2D (left) and 3D (right) Rayleigh-Taylor turbulence. Adapted from Figure~2 of reference \cite{Zhao22JFM}, with permission from Cambridge University Press.
 \label{fig:RT_viz}}
\end{minipage}
\end{figure}

Our method is now applied to a fluid dynamics problem, that of Rayleigh-Taylor (RT) turbulence simulated in both 2D and 3D. RT turbulence occurs when a heavy fluid is supported from below by a light fluid in the presence of gravity or an equivalent acceleration field. An initial small perturbation at the fluid interface grows due to the buoyancy-driven RT instability. It is a fluid dynamics problem that is of both fundamental interest and practical importance in many fields. For example, RT instability plays a leading role in the propagation of the thermonuclear flame front in supernova explosions  \cite{Supernova00}, and  is a main obstacle degrading energy yields in inertial confinement fusion (ICF)  \cite{Betti16}. 

It may be apparent to many fluid dynamics experts that modeling RT unstable 3D systems using 2D simulations is inferior to using 3D simulations. Yet, such propositions remain vague and speculative without clearly identifying the precise aspects of the dynamics that are misrepresented in a 2D simulation. This is necessary to convince and guide practitioners who are focused on global system modeling while facing finite computational resources. For example, ICF modelers aim to predict or postdict results from several shots (\textit{i.e.} experiments) per day in the laboratory. Conducting a single 3D simulation can take several weeks on today's high-end supercomputers, even without incorporating important system components such as the hohlraum and laser physics \cite{Haines2020,GatuJOHNSON2020}. The trade-off ICF modelers face is between including more physical processes in simulations at the expense of conducting them in 2D, despite the recognition that 3D modeling is preferable. To many, trade-offs between 2D and 3D flow physics are not as clear as trade-offs between whether or not to include  components that are critical to the system's evolution such  
laser deposition, hohlraum physics, or resolving small-scale effects such as from the tent or stalk-mount \cite{Clarketal16}. This is why in ICF, 2D simulations are still the main ``work horse'' for experimental design \cite{Schlossberg2021, Kritcher2022-1,Kritcher2022-2} as routine 3D simulations are prohibitively expensive. In the hope of improving 2D modeling, we shall present clear evidence of the stark differences in the fundamental RT scale physics between 2D and 3D models. 

At high Reynolds numbers and at sufficiently late times, the system will develop into turbulence \cite{Zhou17-1,Zhou17-2,Livescu20}. RT turbulence is inhomogeneous, anisotropic, and unsteady. Since RT turbulence is inhomogeneous in the vertical $z$-direction (along gravity), it is a good representative of the challenges faced when analyzing scales in inhomogeneous flows. Previous work had been restricted to analyzing scales in the horizontal directions, which are homogeneous \cite{Zhou17-1, Zhou17-2}. An important exception is the work of Soulard and Griffond \cite{soulard2012inertial} who developed a successful theory for the anisotropic spectral scaling in the inertial range of RT flows. They employed Fourier analysis on the fluctuations in all directions, including the inhomogeneous direction, after removing Reynolds averages. The Soulard-Griffond RT theory applies to scales that are sufficiently small such that the nonlinear term dominates over buoyancy and anisotropy can be regarded as a perturbation around a Kolmogorov-Obukhov equilibrium spectrum \cite{ishihara2002anisotropic,Chertkov03}. Soulard and Griffond~\cite{soulard2012inertial} tested their theory using direct numerical simulations of RT and calculated Fourier transforms of the fluctuating fields along all directions, including the inhomogeneous vertical direction \cite{grea2013rapid}. 

While we agree with \cite{soulard2012inertial} that Fourier analysis can be well-posed for a component of the RT flow (at times when mixing zone is sufficiently small such that the fluctuating fields are at sufficiently small scales relative to the vertical domain size), our goal here is to demonstrate an approach that generalizes Fourier analysis and can be used to diagnose scales of the entire RT flow, including the Reynolds mean fields. As mentioned in the introduction, there is often a lack of scale separation between the mean and fluctuating fields \cite{Zhao22JFM,buzzicotti2023spatio}, necessitating a characterization of spectral information for both. 
Moreover, Fourier analysis, due to its associated Dirichlet kernel that is highly nonlocal in $x$-space (Fig.~\ref{fig:FilterKernels} in Appendix) presents additional difficulties to scale analysis of inhomogeneous flows. Scales have to be limited to sufficiently large Fourier wavenumbers $k$ in the vertical and to sufficiently early times in the RT mixing layer growth to avoid domain boundary artifacts \cite{soulard2012inertial}.

Recently, reference \citep{Zhao22JFM} measured the 1-dimensional filtering spectra (see Fig.~\ref{fig:KE_spec_three_kernels} below) for RT turbulence, including scales in the vertical inhomogeneous direction, but did not distinguish the spectral contribution from different directions. Here, we examine shape anisotropy as a function of scale. In addition, reference \citep{Zhao22JFM} showed the limitation of 2D RT simulations as surrogates for 3D configurations from the perspective of energy transfer across scales. Here we shall extend the comparison between 2D and 3D RT from the perspective of scale-dependent anisotropy.

Visualizations of the simulated flow in 2D and 3D are shown in figure \ref{fig:RT_viz}. They convey qualitatively the turbulent nature of these flows (see \cite{Zhao22JFM}). The simulations are of the fully compressible Navier-Stokes equation with an ideal gas equation of state. The governing equations are solved in 2D and 3D rectangular domains in Cartesian coordinates. The boundary conditions are periodic in the horizontal directions and no-slip rigid walls in the vertical $z$-direction. The equations are solved using a hybrid solver, which uses the pseudo-spectral scheme in the horizontal and a 6th order compact finite difference scheme in the vertical. Here, we use data from simulations 3D1024 and 2D4096 described in Table 1 in \cite{Zhao22JFM}, with the Atwood number 0.5 and the domain-size based Reynolds number 13854 and 44562 for 3D1024 and 2D4096, respectively. More details can be found in \cite{Zhao22JFM}.

As discussed in Appendix section\ref{sec:MarginalSpectra}, extracting the 3-dimensional cumulative spectrum, $\mathcal{E}(\bk)$ in eq.~\eqref{eq:cumu_filtered_energy}, can be computationally expensive. For expediency when analyzing the 3D RT data, we probe the following 2-dimensional scale sub-spaces, $(\ell_x,0,\ell_z)$ and $(\ell_x,\ell_y,0)$, separately. For example, analyzing $(\ell_x,0,\ell_z)$ is accomplished by convolving with kernel $G_{\vell}(\bx)=G_{\ell_x}(x)\,\delta(y)\,G_{\ell_z}(z)$. In this case, no scale decomposition is performed in the $y$-direction, but the convolution is done at every location $(x,y,z)$ in 3D physical space. 

In this section, we use the Gaussian kernel in eq.~\eqref{eq:GaussKernel}. For the definition of filtering wavenumbers, $k=L/\ell$, we use $L=3.2$, which is the  domain size along the horizontal direction.
In simulations with non-periodic boundary conditions, such as our RT flows with rigid walls at the top and bottom,
filtering near the boundary requires a choice for the fields beyond the boundary.
Following past work \citep{Aluieetal2018_jpo,zhao2018inviscid,Zhao22JFM}, we extend the domain beyond the physical boundaries in a manner consistent with the boundary 
conditions. For our RT problem, velocity is kept zero beyond the walls, density field is kept constant (zero normal gradient), and the extended pressure field
satisfies the hydrostatic condition $dP/dz=-\rho g$. The coarse-graining operation~\eqref{eq:filtering} can then be performed at every point in the flow domain.

\subsubsection{3D RT\lb{sec:3DRT}}
\begin{figure}[bhp]
\centering 
\begin{minipage}[b]{1.0\textwidth}  
\centering
\subfigure[~cumulative spectrum $\mE(k_x,\infty,k_z)$]
{\includegraphics[width=0.48\textwidth]{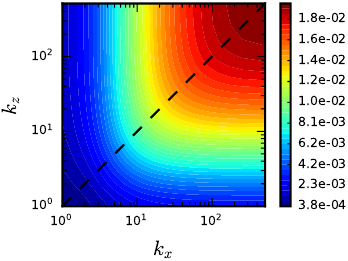} } 
\subfigure[~2D filtering spectrum $\OL{E}^{2\textrm{D}}(k_x,k_z)$]
{\includegraphics[width=0.48\textwidth]{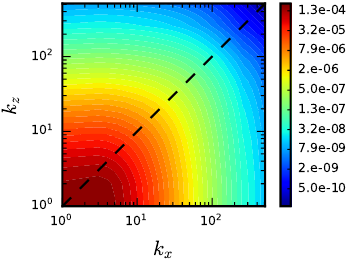}} \\
\subfigure[~scale-dependent centroid]
{\includegraphics[width=0.42\textwidth]{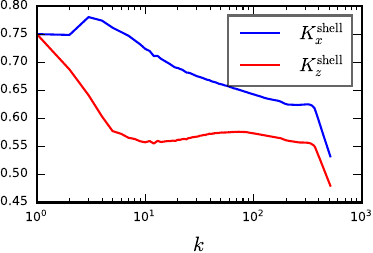} }
\subfigure[~scale-dependent anisotropy metric]
{\includegraphics[width=0.42\textwidth]{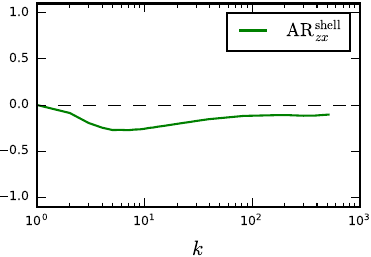} }
    \caption{3D-RT filtering spectra of KE along the $x$-$z$ directions using Gaussian kernels (eq.~\eqref{eq:GaussKernel}). Panel~(a) is the cumulative spectrum (eq.~\eqref{eq:cumu_filtered_energy}) in filtering wavenumber subspace $(k_x,k_z)$. Panel~(b) is the associated 2D filtering spectrum obtained via eq.~\eqref{eq:anisotropic_spectra} (note the logarithmic color bar). Panel~(c) shows the scale-dependent centroid obtained from integrating $\OL{E}^{2\textrm{D}}$ within a shell (eq.~\eqref{eq:integral_K_ann}). Panel~(d) shows the anisotropy metric from eq.~\eqref{eq:AR_ann}. Note the asymmetry relative to the diagonal in panels~(a),(b), associated with vertically elongated flow structures at large scales (small $k$). The asymmetry seems to become less at smaller scales (larger $k$), which is made clear in panel~(c) where the centroid moves closer to the diagonal at high $k$  ($K_x\approx K_z$). This is also reflected in the anisotropy metric in~(d), where $AR\to 0$ as $k\to\infty$.
 \label{fig:3D_RT_KE_xz}}
\end{minipage}
\end{figure}

\begin{figure}[bhp]
\centering 
\begin{minipage}[b]{0.45\textwidth}  
\raggedright
\subfigure[~coarse density $\OL{\rho}_\ell$~~~~~~~~~~~~~~~~~~~]
{\includegraphics[height=0.95\textwidth]{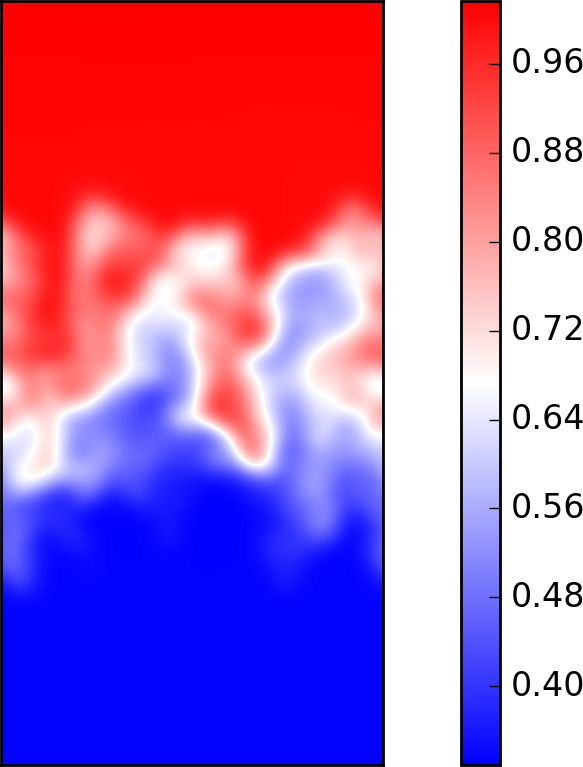} }
\end{minipage}
\begin{minipage}[b]{0.45\textwidth}  
\raggedright
\subfigure[~coarse KE $\mE$~~~~~~~~~~~~~~~~~~~]
{\includegraphics[height=0.95\textwidth]{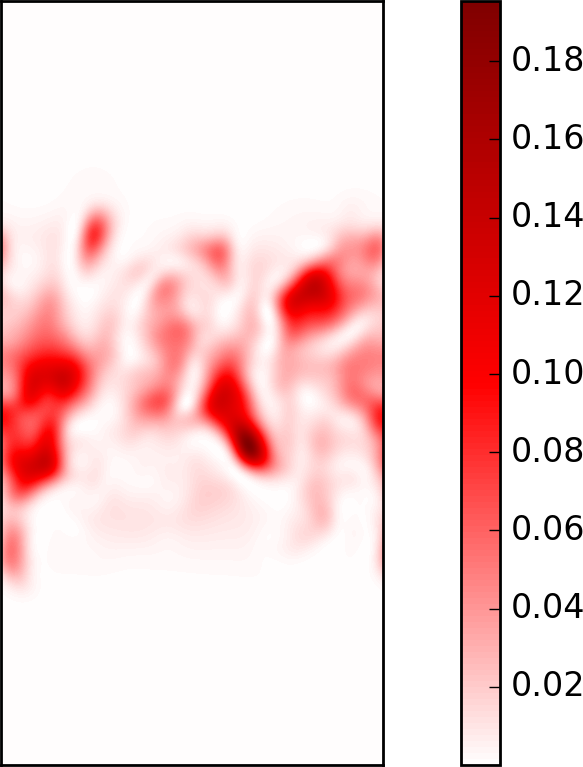}}
\end{minipage} \\[12pt]
\begin{minipage}[b]{0.45\textwidth}  
\raggedright
\subfigure[~coarse streamlines]
{\includegraphics[height=0.95\textwidth]{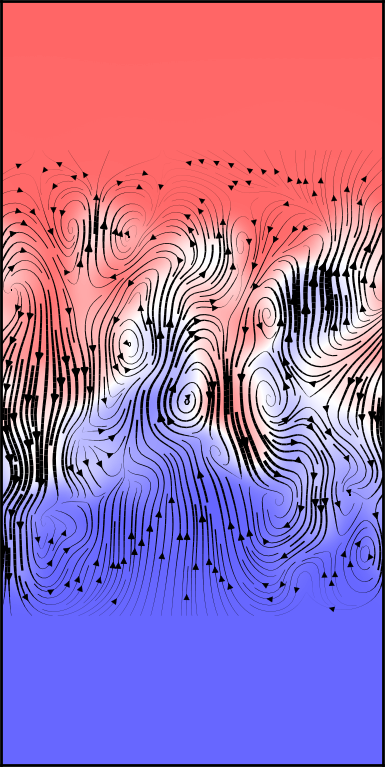} } \end{minipage}
\begin{minipage}[b]{0.45\textwidth}  
\raggedright
\subfigure[~fine KE $\mE_{<\ell}$~~~~~~~~~~~~~~~~~~~]
{\includegraphics[height=0.95\textwidth]{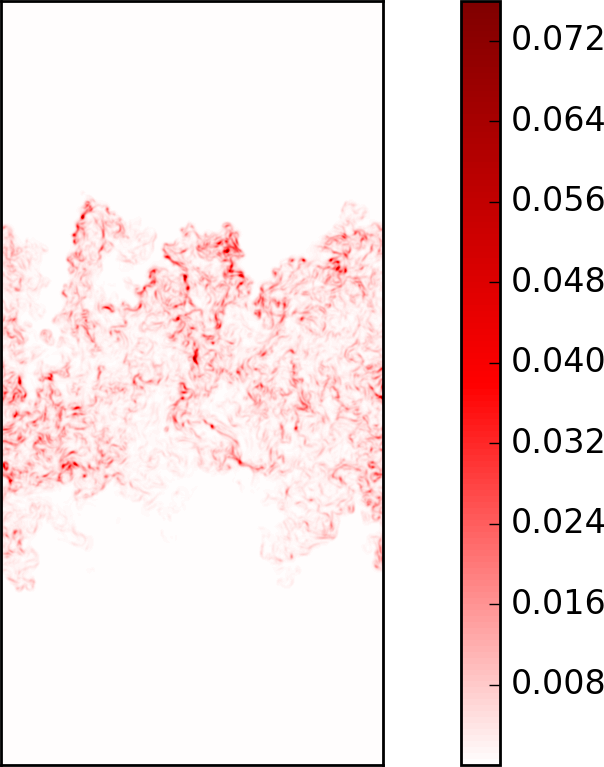} }
\end{minipage} 
    \caption{3D-RT at different scales. Visualizations show a 2D slice at the mid-plane $y=L/2$, where $L$ is the domain extent in the horizontal direction. Panels~(a),(b) show the coarse-grained density, $\OL{\rho}_\ell$ in (a), and kinetic energy, $\mE(k_x,k_z)$ in (b), at scale $k_x=k_z=L/\ell=10$. The filtering kernel is a Gaussian (eq.~\eqref{eq:GaussKernel}) with $\ell_x=\ell_z=\ell$.
    Panel~(c) shows $\OL\rho_\ell$ from panel (a) in the background and the streamlines based on the coarse-grained velocity field $\left(\OL{\left(\rho\,u_x\right)}_\ell/\OL\rho_\ell,\OL{\left(\rho\,u_z\right)}_\ell/\OL\rho_\ell\right)$ with filtering as in panels (a),(b) at $k=10$. Panel~(d) shows the small-scale (or fine) KE, $\mE_{<\vell}$ defined in eq.~\eqref{eq:fineKE}, with $k_x=k_z=100$. Anisotropy at large scales is evident in panels~(a)-(c) while small scales seem to regain a semblance of isotropy.
 \label{fig:3D_RT_rho_KE_filter_viz}}
\end{figure}

To quantify scale anisotropy in 3D RT, we first calculate $\mathcal{E}(k_x,\infty,k_z)$ and $\OL{E}^{2\textrm{D}}(k_x,k_z)$ in filtering wavenumber subspace $(k_x,k_z)$. These are shown in Fig.~\ref{fig:3D_RT_KE_xz}. The asymmetry relative to the diagonal $k_z=k_x$ is especially clear from $\OL{E}^{2\textrm{D}}$ in Fig.~\ref{fig:3D_RT_KE_xz}b at wavenumbers smaller than $\approx 20$, where $\OL{E}^{2\textrm{D}}(k_x,k_z)$ is skewed toward smaller $k_z$ (\textit{i.e.} larger vertical length scales $\ell_z\sim k_z^{-1}$). This indicates that the flow is elongated in the vertical, which is consistent with the notion of large scale vertically inter-penetrating bubbles and spikes. Further evidence for large scale anisotropy is derived from Fig.~\ref{fig:3D_RT_rho_KE_filter_viz}, where vertically elongated shapes are visually obvious in panels (a)-(c) showing the flow at scales larger than $\ell=L/10$. Fig.~\ref{fig:3D_RT_rho_KE_filter_viz} underscores the utility of the coarse-graining approach to analyzing anisotropy in a natural and intuitive manner.

The asymmetry relative to the diagonal is less clear at higher wavenumbers in Fig.~\ref{fig:3D_RT_KE_xz}b. Indeed, visualizing the corresponding small-scales in physical space in Fig.~\ref{fig:3D_RT_rho_KE_filter_viz}d suggests that they are in approximate isotropy. The quantity shown in Fig.~\ref{fig:3D_RT_rho_KE_filter_viz}d is energy at scales smaller than $\ell$ and defined as
\be \mE_{<\vell} =  \frac{1}{2}\left\{ \OL{\left(\rho|\bu|^2\right)}_{\vell}\,-\frac{|\OL{\rho\bu}_{\vell}|^2}{\OL{\rho}_{\vell}}\right\}~,
\lb{eq:fineKE}\ee
which is positive semi-definite as proved in \cite{Sadek18}.

Panels (c) and (d) in Fig.~\ref{fig:3D_RT_KE_xz} provide a more quantitative measure of anisotropy. Discrepancy between $K_x^{\textrm{shell}}$ and $K_z^{\textrm{shell}}$ is clear around $k=5$, where their ratio reaches $\approx 4/3$. This discrepancy is better captured by the anisotropy metric $AR_{zx}^{\textrm{shell}}(k)$ in Fig.~\ref{fig:3D_RT_KE_xz}(d), which has an extremum of $AR_{zx}^{\textrm{shell}}(5)=-0.27$. $AR_{zx}^{\textrm{shell}}(k)$ trends toward zero at higher $k$, which indicates approximate isotropy field at smaller length scales, although it never reaches zero over the dynamic range of scales we resolve in the simulation. This is consistent with the presence of residual small-scale \emph{vector} anisotropy observed in Livescu \textit{et al.} \cite{Livescu09} (Fig.~14 in \cite{Livescu09}). It is also consistent with anisotropic (or ``directional'') spectra of the velocity fluctuations in Soulard and Griffond \cite{soulard2012inertial} (Figs.~6-7 in \cite{soulard2012inertial}). We are unsure if this residual anisotropy at high $k$ in Fig.~\ref{fig:3D_RT_KE_xz}(c),(d) would persist in the limit of higher Reynolds numbers. Appendix Figs.~\ref{fig:1st_3rd_kernel}-\ref{fig:1st_3rd_anisotropy_metrics} repeat the anisotropy analysis using a Boxcar and a 3rd-order filtering kernel, which in some sense is more similar to a truncated Dirichlet kernel \cite{spyropoulos1994evaluation}, showing that our results are insensitive to the Gaussian kernel we are using.

Panels~\ref{fig:3D_RT_KE_xz}(c),(d) show that at scales $\ell$ approaching the horizontal domain size $L$ ($k=1$), we have $K_x^{\textrm{shell}}(1)=K_z^{\textrm{shell}}(1)$ and $AR_{zx}^{\textrm{shell}}(1)\approx 0$ indicating isotropy. At first glance, this may seem surprising since we expect RT flows to be highly anisotropic at large scales \cite{Livescu09,soulard2012inertial,Zhou17-1}. However, note that scale $\ell=L$ ($k=1$) is much larger than the largest bubble or spike in Fig.~\ref{fig:RT_viz}. As the coarse-graining scale $\vell$ approaches that of the domain size, spatially local KE, $\sfrac{1}{2}|\OL{\rho\bu}_\vell|^2/\OL\rho_\vell$,  approaches that of a domain average, $\sfrac{1}{2}|\{\rho\bu\}|^2/\{\rho\}$. A domain average has trivial shape isotropy because it is spatially constant. We saw this in our illustrative example of an anisotropic Gaussian in Fig.~\ref{fig:synthetic_data_viz}(a), which is isotropic at scales comparable to the domain size as shown in Fig.~\ref{fig:Gaussian_filtered_Gaussian_viz}. For 3D-RT, we have a similar situation in Fig.~\ref{fig:3D_RT_KE_xz}(d) where $AR^\mathrm{shell}(k=1)\approx0$. This can be seen from the visualization of coarse KE at $k=1$ in Fig.~\ref{fig:filtered_KE_viz} of the Appendix.

\begin{figure}[bhp]
\centering 
\begin{minipage}[b]{1.0\textwidth}  
\centering
\subfigure[~cumulative spectrum $\mE(k_x,k_y,\infty)$]
{\includegraphics[width=0.48\textwidth]{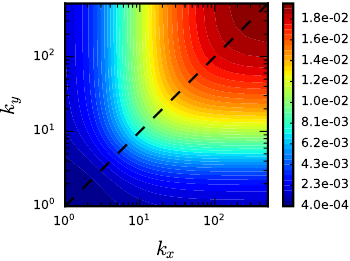} } 
\subfigure[~2D filtering spectrum $\OL{E}^{2\textrm{D}}(k_x,k_z)$]
{\includegraphics[width=0.48\textwidth]{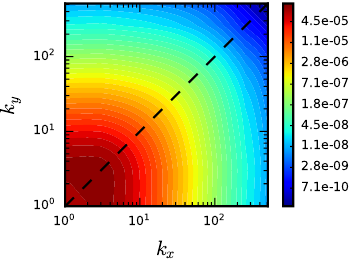}} \\
\subfigure[~scale-dependent centroid]
{\includegraphics[width=0.42\textwidth]{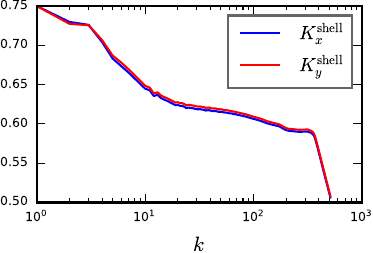} }
\subfigure[~scale-dependent anisotropy metric]
{\includegraphics[width=0.42\textwidth]{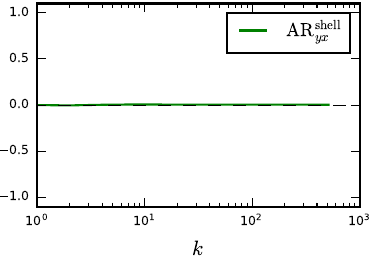} }
    \caption{Same as in Fig.~\ref{fig:3D_RT_KE_xz} but in the horizontal $x$-$y$ directions for 3D-RT. The flow is isotropic at all scales as expected.
 \label{fig:3D_RT_KE_xy}}
\end{minipage}
\end{figure}

In contrast to Fig.~\ref{fig:3D_RT_KE_xz}, comparing scales in the two horizontal ($x$-$y$) directions in Fig.~\ref{fig:3D_RT_KE_xy} shows almost perfect isotropy, consistent with physical expectations. We see that $\mathcal{E}(k_x,k_y,\infty)$ and $\OL{E}^{2\textrm{D}}(k_x,k_y)$ are symmetric relative to the diagonal $k_y=k_x$, and $AR_{yx}^{\textrm{shell}}=0$ for all $k$.

\clearpage
\subsubsection{2D RT}
\begin{figure}[bhp]
\centering 
\begin{minipage}[b]{1.0\textwidth}  
\centering
\subfigure[~cumulative spectrum $\mE(k_x,k_z)$]
{\includegraphics[width=0.48\textwidth]{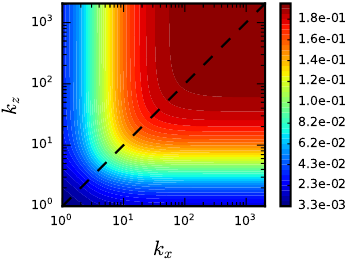} } 
\subfigure[~2D filtering spectrum $\OL{E}^{2\textrm{D}}(k_x,k_z)$]
{\includegraphics[width=0.48\textwidth]{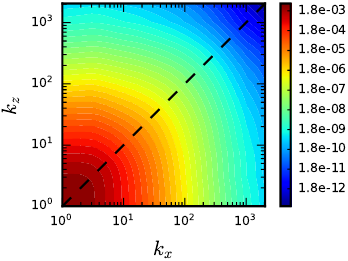}} \\
\subfigure[~scale-dependent centroid]
{\includegraphics[width=0.42\textwidth]{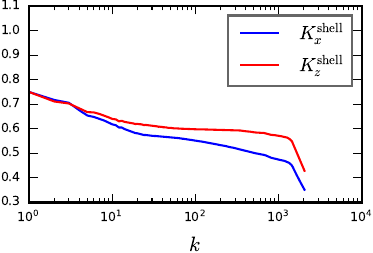} }
\subfigure[~scale-dependent anisotropy metric]
{\includegraphics[width=0.42\textwidth]{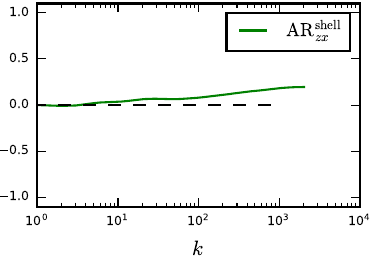} }
    \caption{2D-RT filtering spectra of KE as in Fig.~\ref{fig:3D_RT_KE_xz}. Note in panels~(a),(b) the symmetry relative to the diagonal at large-scales (small $k$), associated with a quasi-isotropic overturning circulation. However, asymmetry emerges at smaller scales (larger $k$), which is made clear in panel~(c) where the centroid moves away from the diagonal at high $k$  ($K_x<K_z$), associated with horizontal layering due to stable stratification. This is also reflected in the anisotropy metric in~(d), where $AR > 0$ as $k\to\infty$.
 \label{fig:2D_RT_KE_xz}}
\end{minipage}
\end{figure}

To quantify scale anisotropy in 2D RT, we calculate $\mathcal{E}(k_x,\infty,k_z)$ and $\OL{E}^{2\textrm{D}}(k_x,k_z)$, shown in Fig.~\ref{fig:2D_RT_KE_xz}. From panels (a) and (b), the asymmetry relative to the diagonal $k_z=k_x$ seems absent at small filtering wavenumbers but becomes discernible at $k\gtrsim 50$, where $\OL{E}^{2\textrm{D}}(k_x,k_z)$ is skewed slightly toward smaller $k_x$ (\textit{i.e.} larger horizontal length scales $\ell_x\sim k_x^{-1}$). This indicates that the flow is elongated in the horizontal, which may be surprising for Rayleigh-Taylor flows where it is expected that the flow would be elongated in the vertical due to buoyancy forcing \cite{Livescu08,Soulard12,Zhou17-2}, similar to what we observed in section~\ref{sec:3DRT} in 3D-RT.

Panels (c) and (d) in Fig.~\ref{fig:2D_RT_KE_xz} provide a more quantitative measure of anisotropy. We see that $K_x^{\textrm{shell}}(1)=K_z^{\textrm{shell}}(1)$ at the smallest wavenumbers. The discrepancy between $K_x^{\textrm{shell}}$ and $K_z^{\textrm{shell}}$ becomes discernible at $k>10$. This is also captured by $AR_{zx}^{\textrm{shell}}(k)$ in Fig.~\ref{fig:2D_RT_KE_xz}(d), which is positive and increases at larger $k$, indicating that the flow is elongated in the horizontal and becomes increasingly anisotropic at those smaller length scales. 

\begin{figure}[bhp]
\centering 
\begin{minipage}[b]{0.45\textwidth}  
\raggedright
\subfigure[~coarse density $\OL{\rho}_\ell$~~~~~~~~~~~~~~~~~~~]
{\includegraphics[height=0.95\textwidth]{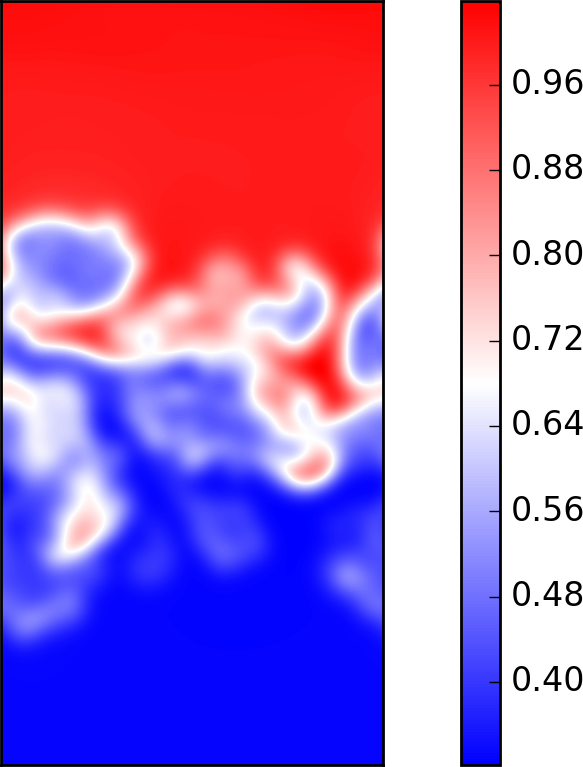} } 
\end{minipage}
\begin{minipage}[b]{0.45\textwidth}  
\raggedright
\subfigure[~coarse KE $\mE$~~~~~~~~~~~~~~~~~~~]
{\includegraphics[height=0.95\textwidth]{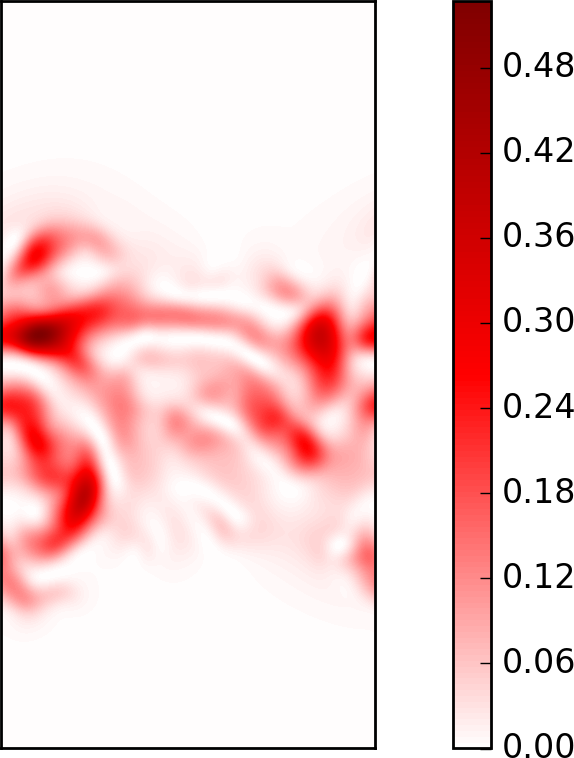}} 
\end{minipage} \\
\begin{minipage}[b]{0.45\textwidth}  
\raggedright
\subfigure[~coarse streamlines]
{\includegraphics[height=0.95\textwidth]{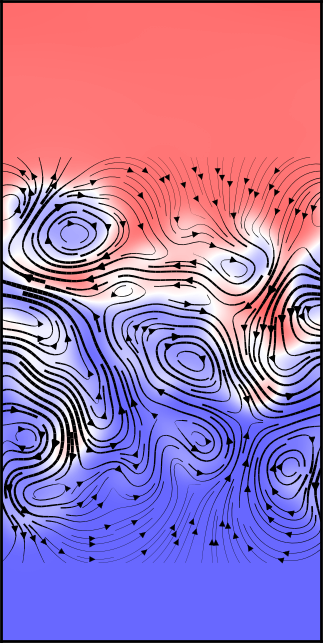} }
\end{minipage}
\begin{minipage}[b]{0.45\textwidth}  
\raggedright
\subfigure[~fine KE $\mE_{<\ell}$~~~~~~~~~~~~~~~~~~~]
{\includegraphics[height=0.95\textwidth]{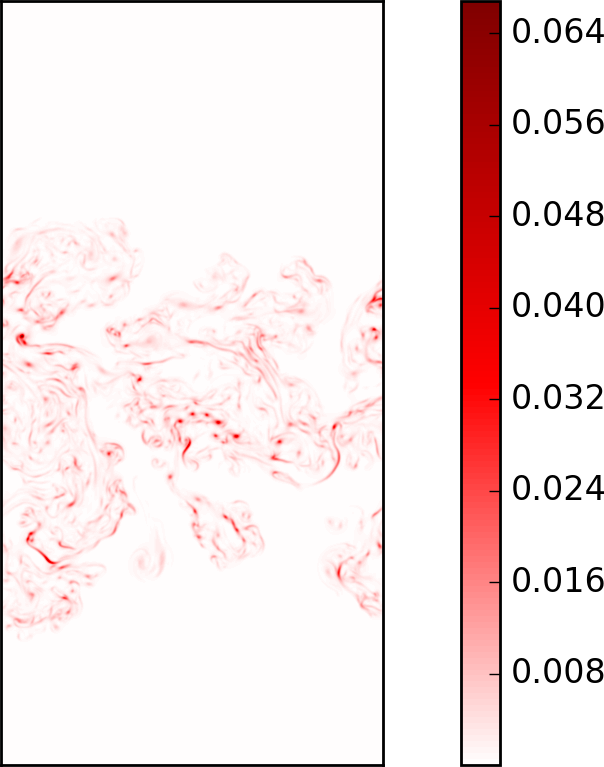} } 
\end{minipage}
    \caption{2D-RT at different scales, as in Fig.~\ref{fig:3D_RT_rho_KE_filter_viz}. Panels~(a)-(c) are filtered at scale $k_x=k_z=L/\ell=10$ using a Gaussian kernel. Panel~(d) shows small-scale KE at $k_x=k_z=100$.  Large-scale flow in panels~(a)-(c) is approximately isotropic, which is especially clear in the large-scale overturning circulation in panel~(c). In contrast, the small-scale flow in (d) shows a tendency for horizontal layering due to the stable stratification that results from overturning.
 \label{fig:2D_RT_rho_KE_filter_viz}}
\end{figure}

\begin{figure}[bhp]
\centering 
\begin{minipage}[b]{1.0\textwidth}  
\centering
\subfigure[~coarse density $\OL{\rho}_\ell$~~~~~~~~~~~~~~~~~~~]
{\includegraphics[height=0.45\textwidth]{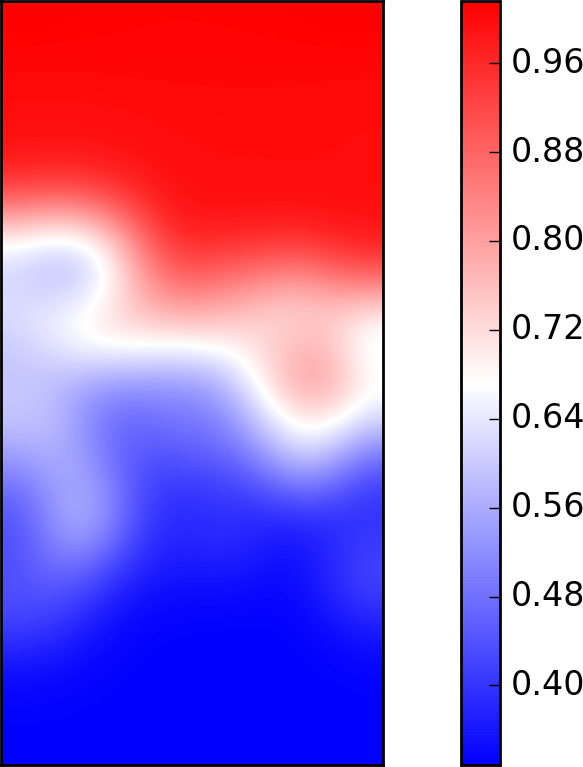} }
\subfigure[~coarse KE $\mE$~~~~~~~~~~~~~~~~~~~]
{\includegraphics[height=0.45\textwidth]{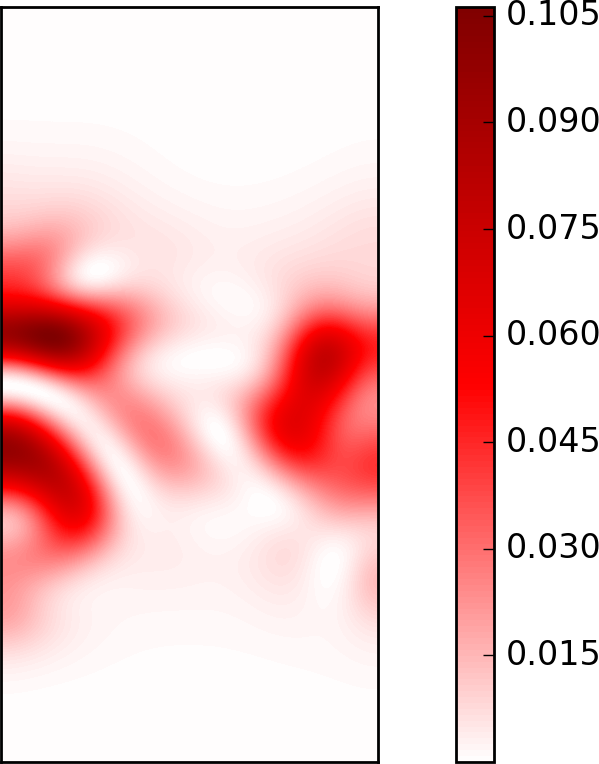}} \phantom{pp}
\subfigure[~coarse streamlines]
{\includegraphics[height=0.45\textwidth]{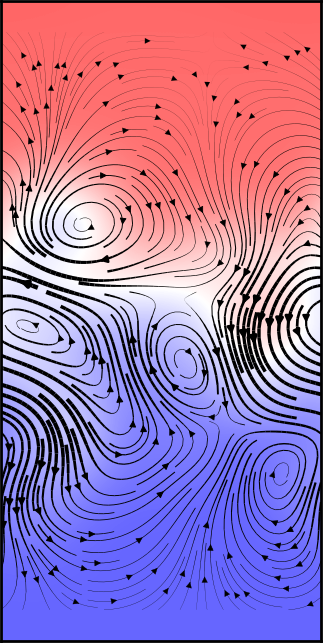} }
    \caption{2D-RT has a quasi-isotropic large-scale overturning circulation due to an upscale cascade~\cite{Zhao22JFM}, which is absent in 3D-RT. Panels~(a)-(b) are similar to those in Fig.~\ref{fig:2D_RT_rho_KE_filter_viz} but filtered at scale $k_x=k_z=L/\ell=3$. Panel~(c) demonstrates clearly the large-scale overturning circulation.
 \label{fig:2D_RT_rho_KE_filter_k3_viz}}
\end{minipage}
\end{figure}

Isotropy at large scales and increasing anisotropy at small scales in 2D RT is consistent with recent results in \cite{Zhao22JFM}. In that work, it was shown that 2D RT is characterized by the emergence of a large scale overturning circulation, which isotropizes the 2D flow. This is in contrast to the picture of vertically rising bubbles and sinking spikes often invoked for RT flows, which nevertheless seems consistent with 3D-RT. 
Direct evidence for the existence of a large scale overturning circulation in 2D RT is provided in Figs.~\ref{fig:2D_RT_rho_KE_filter_viz},\ref{fig:2D_RT_rho_KE_filter_k3_viz}. The overturning circulation in Fig.~\ref{fig:2D_RT_rho_KE_filter_viz}c is quasi-isotropic and is absent at the same large scales ($>\ell=L/10$) in 3D RT shown in Fig.~\ref{fig:3D_RT_rho_KE_filter_viz}c, which are clearly anisotropic. Fig.~\ref{fig:2D_RT_rho_KE_filter_k3_viz} shows even larger scales ($>\ell=L/3$) in 2D RT, where the large-scale isotropic overturning circulation is clearer. In the Appendix Fig.~\ref{fig:regional_aniso_metrics}, we calculate the anisotropy metrics at different locations along the vertical direction.

It was shown in \cite{Zhao22JFM} that 2D RT does not cascade energy to smaller scales, resulting in weaker small inertial scales relative to 3D-RT. Weaker inertial scales and an associated suppressed molecular mixing was also discussed in several previous studies \cite{Cabot06POF,Zhou17-1,Huasen19,Bian20}. 

Therefore, our results in Fig.~\ref{fig:2D_RT_KE_xz} indicate that the combination of a large-scale overturning and suppressed small-scale mixing in 2D-RT yields strong stable stratification at small-scales. Such stratification leads to a flow that is elongated in the horizontal as quantified by the anisotropy metrics in Fig.~\ref{fig:2D_RT_KE_xz}c,d. Direct visual evidence for the anisotropic flow at small scales is provided in Fig.~\ref{fig:2D_RT_rho_KE_filter_viz}d, which should be contrasted to the corresponding Fig.~\ref{fig:3D_RT_rho_KE_filter_viz}d in 3D RT.

\clearpage
\section{Conclusions \label{sec:conclude}}
This work establishes how spatial coarse-graining can be used to quantify anisotropy at different length scales. It is based on generalization of ideas from \cite{Sadek18}, which we demonstrated via simple illustrative examples and an application to Rayleigh-Taylor turbulence.

In addition to the method, a main result of this work showed that anisotropy of Rayleigh-Taylor turbulence in 2D is fundamentally different from that of 3D-RT. We showed that 2D-RT is characterized by enhanced anisotropy at small scales in a manner that was perhaps unexpected, but in hindsight is consistent with recent work \cite{Zhao22JFM}. A clear identification of how 2D modeling of applications involving RT can misrepresent the actual 3D hydrodynamics is a necessary step to develop new modeling strategies without having to conduct fully 3D simulations. This is because 3D hydrodynamics modeling remains prohibitively expensive in many applications such as ICF as discussed above.

We envisage that our method for quantifying shape anisotropy would be especially valuable in flows that are inhomogeneous, where established methods such as Fourier analysis cannot be used in a straightforward manner. In this regard, it is important to be aware of limitations of filtering spectra. In particular, it was shown in \cite{Sadek18} that when using first-order kernels such as a Gaussian, filtering spectra do not decay faster than $k^{-3}$ as $k\to\infty$, even if there is little or no energy at those small scales when probed with more complex (higher-order) kernels. Therefore, determining whether or not energy conveyed by filtering spectra is physical or a spurious artifact of the kernel being used requires using higher order kernels if a $k^{-3}$ scaling is observed. Regarding anisotropy, the potential for `locking' in at a $k^{-3}$ scaling when using first-order filtering kernels would still allow us to detect anisotropy using the metrics, $\bK^\mathrm{shell}(k)$ and $\mathrm{AR}_{ij}^\mathrm{shell}(k)$, although the level of anisotropy may be underestimated.

\begin{acknowledgments}
We thank Olivier Soulard and two anonymous reviewers for their valuable comments that helped improve our paper.
This research was funded by US DOE grant DE-SC0020229 and NSF grant PHY-2206380. Partial support from US NSF grants PHY-2020249 is acknowledged. HA was also supported by US DOE grants DE-SC0014318, DE-SC0019329, US NSF grant OCE-2123496, US NASA grant 80NSSC18K0772, and US NNSA grants DE-NA0003856, DE-NA0003914, DE-NA0004134. DZ also acknowledges financial support from the National Natural Science Foundation of China (No. 12202270).  Computing time was provided by NERSC under Contract No. DE-AC02-05CH11231. 
\end{acknowledgments}

\clearpage

\renewcommand{\theequation}{A-\arabic{equation}}
\renewcommand{\thefigure}{A\arabic{figure}}
\renewcommand{\thesection}{\Alph{section}} 
\renewcommand{\thesubsection}{\Alph{subsection}}
\setcounter{equation}{0}  
\setcounter{figure}{0}  
\setcounter{Prop}{0}  
\setcounter{section}{0}  
\section*{APPENDIX}  

\subsection{Filtering kernels and Fourier analysis}
\begin{figure}[h]
\centering
\includegraphics[width=.4\textwidth]{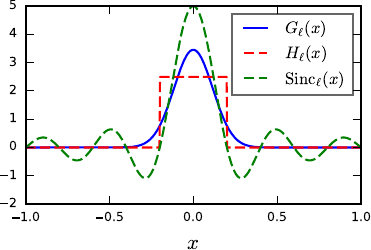}
\includegraphics[width=.4\textwidth]{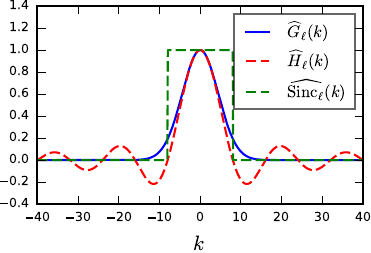}
	\caption{\footnotesize{(left) Comparing the Gaussian ($G_\ell$), Boxcar ($H_\ell$), and Dirichlet ($\mbox{Sinc}_\ell$) kernels in physical space, with the filter width $\ell=L_x/8=0.4$.} The Boxcar is exactly zero beyond a distance $\ell/2$ from the center, the Gaussian decays faster than exponentially, while the Dirichlet is highly non-local and decays as $1/|x-x_0|$ away from the center $x_0=0$. (right) The same kernels shown in Fourier $k$-space.}
	\label{fig:FilterKernels}
\end{figure}

\lb{sec:FilteringKernels}
Fig.~\ref{fig:FilterKernels} shows three example kernels that can be used to calculate the filtering spectrum. When using the Dirichlet kernel, the filtering spectrum is exactly the same as the Fourier spectrum. The Dirichlet is highly non-local and decays as $1/|x-x_0|$ away from the center at $x_0$, where the filtered field $\OL{u}(x_0)$ is being evaluated. Therefore, the filtered field at location $x_0$ has contributions from spatial locations $x$ far away from $x_0$. This is in comparison to filtering with a more localized kernel, such as a Gaussian or Boxcar, which is practically (or exactly) zero beyond a distance $\sim\ell/2$ away from $x_0$. 

Therefore, in a bounded domain, regions outside the boundaries can make non-negligible (possibly spurious) contributions to the Fourier scale analysis (equivalent to filtering with a Dirichlet kernel) even in regions that are presumably far from the boundaries. When using a more compact kernel, such as the Gaussian or Boxcar, the ``domain of dependence'' is localized to be within $\sim\ell/2$ from the location being analyzed. This makes it feasible to extend the fields beyond the domain boundaries (in a manner satisfying the boundary conditions) to analyze scales near those boundaries.

\subsection{Marginal Filtering Spectra}\lb{sec:MarginalSpectra}
The cumulative spectrum, $\mE$, or equivalently, the filtering spectrum, $\OL{E}^{nD}$, evaluated over the entire scale-grid (Fig.~\ref{fig:3DfilteringSpec}) contains all information about the energy content of spatial scales in different directions. However, $\mE$ can be computationally expensive to calculate using data on $N^n$ grid-points in $n$-dimensions   since it relies on spatial filtering via convolutions. Scanning the entire range of scales in each direction requires nominally $O(N^n)$ different scale vectors $(\ell_1,\dots,\ell_n)$, each of which can cost of $O(N^{2n})$ operations for the convolution. Therefore, in 3-dimensions, the cost can be as high as $O(N^{9})$. While there are several algorithmic optimizations to reduce the cost \citep{StorerAluieJOSS,Storer2022NatComm,buzzicotti2023spatio}, such as subsampling the scale-grid (especially at high $k$), the computational cost can still be considerable.
\begin{figure}
\centering
\begin{minipage}[b]{1.0\textwidth}
\centering
    \subfigure[Marginal spectrum in eq.~\eqref{eq:marginal_spectra}]
{\includegraphics[height=0.42\textwidth]{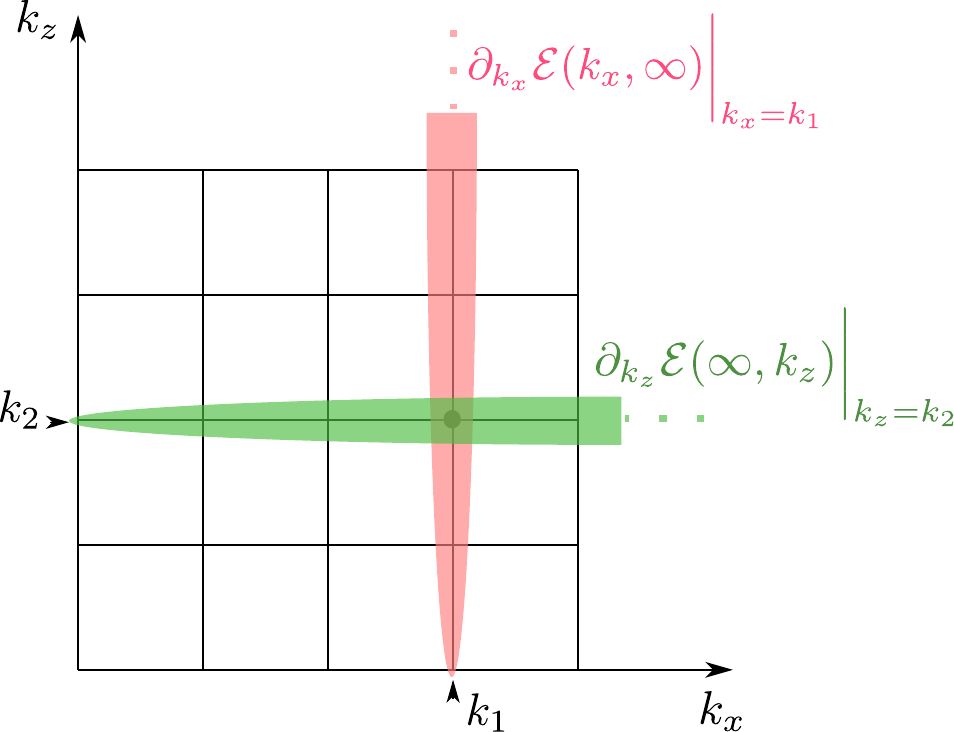}} 
    \phantom{ppp}
     \subfigure[Anisotropy based on marginal centroid]{\includegraphics[width=0.4\textwidth]{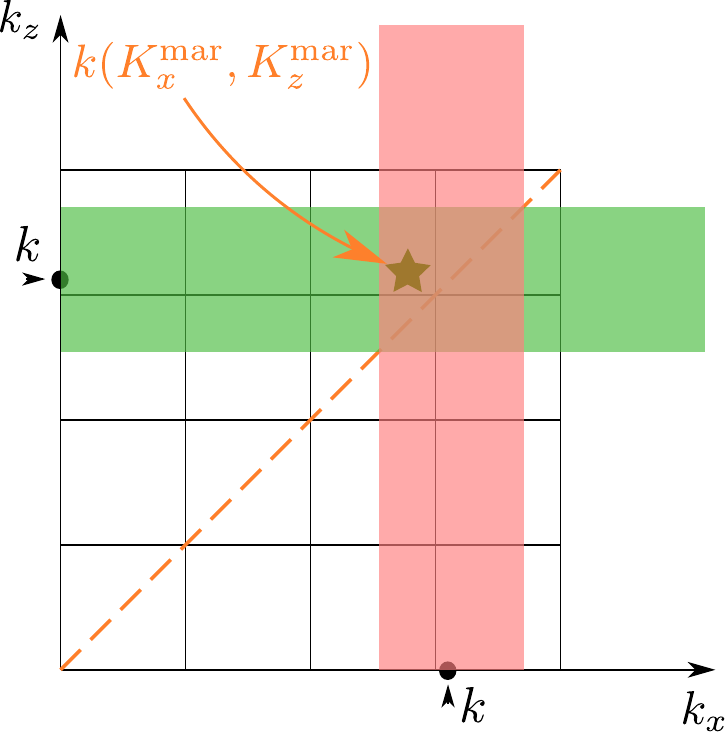}} 
    \caption{
    {\textbf{Panel (a):}} similar to Fig.~\ref{fig:AnisotropySpectraSchema_shell}(a) but for ``marginal spectra'',  $\partial_{k_x}\mE$ and $\partial_{k_z}\mE$. We see that $\partial_{k_x}\mE(k_1,\infty)$ (in red) accounts for field variations at scale $\ell_1=L/k_1$ in the $x$-direction, but contains all scales in the $z$-direction. Similarly, $\partial_{k_z}\mE$ (in green) accounts for variations in the $z$-direction.
{\textbf{Panel (b):}}
    (b) similar to Fig.~\ref{fig:AnisotropySpectraSchema_shell}(b) but for the marginal centroid ($\bK^\mathrm{mar}(k)$ in eq.~\eqref{eq:integral_K_mar}), depicted as a star. It is based on $\OL{E}^{\mathrm{mar}}(k)$, which is more easily computed than $\OL{E}^{\mathrm{shell}}(k)$ but at the expense of accuracy since it incorporates \emph{all} scales in other directions. The restriction on $\bK^\mathrm{mar}(k)$ to reside where the two (red and green) strips overlap also makes it a less sensitive gauge of anisotropy.
\label{fig:AnisotropySpectraSchema_marginal}
}
\end{minipage}
\end{figure}

It is possible to quantify anisotropy by sampling a much smaller subset of the scale-grid but at the cost of accuracy and sensitivity. 
Fig.~\ref{fig:AnisotropySpectraSchema_marginal}(a) sketches what we call the ``marginal spectra,'' in analogy to the marginal probability distribution. A scale decomposition via filtering is performed along each direction separately rather than concurrently. The marginal spectra along different directions are defined as
\begin{subequations}
\label{eq:marginal_spectra}
\begin{align}
\OL{E}_x^\mathrm{mar}(k) &\equiv \frac{\partial}{\partial k_x}\mathcal{E}(k_x, \infty, \infty)\Big\rvert_{k_x=k},\\ 
\OL{E}_y^\mathrm{mar}(k) &\equiv \frac{\partial}{\partial k_y}\mathcal{E}(\infty, k_y, \infty)\Big\rvert_{k_x=k},\\ 
\OL{E}_z^\mathrm{mar}(k) &\equiv \frac{\partial}{\partial k_z}\mathcal{E}(\infty,\infty, k_z)\Big\rvert_{k_z=k} ~.
\end{align}
\end{subequations}
Here, $\infty$ denotes the absence of filtering ($\ell=L/k=0$)  along the corresponding direction, thereby retaining the contribution from all scales along that direction. The marginal spectrum $\OL{E}_x^\mathrm{mar}(k)$, for example, is obtained from $\OL{\bu}_{\ell_x, 0, 0}(\bx)$ by convolving with kernel $G_{\ell_x}(r_x)\delta(r_y)\delta(r_z)$ in eq.~\eqref{eq:AnisoFiltering}, where the delta function $\delta(r) = G_{\ell=0}(r)$. 

Comparing $E_x^\mathrm{mar}(k)$, $E_y^\mathrm{mar}(k)$, and $E_z^\mathrm{mar}(k)$ gives us a gauge of anisotropy at different scales. However, it is important to be mindful of what information is conveyed in eq.~\eqref{eq:marginal_spectra} and its limitations. When integrated over $k$, all three quantities yield the same total energy. Therefore, if $E_z^\mathrm{mar}(k_1) > E_x^\mathrm{mar}(k_1)$ at $k_1$, it is guaranteed that $E_x^\mathrm{mar}(k_2) > E_z^\mathrm{mar}(k_2)$ for some other $k_2$. In comparison, $\OL{E}^{n\mathrm{D}}$ has no similar constraint. 

Similar to eq.~\eqref{eq:integral_K_ann}, the normalized scale-dependent centroid of marginal spectra within 1D shells are
\begin{align} \label{eq:integral_K_mar}
    {K}^\mathrm{mar}_i(k)\equiv
    \begin{cases}
        \left.\displaystyle\frac{1}{k}\int\limits_{k/\sqrt{2}}^{\sqrt{2}k} dq \, q \,\OL{E}_i^\mathrm{mar}(q) \middle/ \displaystyle\int\limits_{k/\sqrt{2}}^{\sqrt{2}k} dq \,  \OL{E}_i^\mathrm{mar}(q), \right.\hspace{1cm}\mathrm{for~}i=1,\dots,n,\\\\
         \hspace{.5cm}0 \hspace{1cm}\text{if $\displaystyle\int\limits_{k/\sqrt{2}}^{\sqrt{2}k} dq \,  \OL{E}_i^\mathrm{mar}(q) < \epsilon^{\textrm{numeric}}$.}
    \end{cases} 
\end{align}
We set $\epsilon^{\textrm{numeric}}=10^{-14}$ below.
Since $\OL{E}_i^\mathrm{mar}$ incorporates \emph{all} scales in other directions as shown in Fig.~\ref{fig:AnisotropySpectraSchema_marginal}(a), the marginal centroid, ${\bK}^\mathrm{mar}$, is a less accurate gauge of anisotropy at any one scale. Moreover, its restriction to reside where the two (red and green) strips overlap in Fig.~\ref{fig:AnisotropySpectraSchema_marginal}(b) also makes it a less sensitive gauge of anisotropy.

Similar to eq.~\eqref{eq:AR_ann}, we define the anisotropy metric associated with marginal spectra as
\begin{align} \label{eq:AR_mar}
\mathrm{AR}_{ij}^\mathrm{mar}(k) \equiv \frac{({K}_i^\mathrm{mar})^2-({K}_j^\mathrm{mar})^2}{({K}_i^\mathrm{mar})^2+({K}_j^\mathrm{mar})^2}, \hspace{1cm}\mathrm{for~}i,j=1,\dots,n.
\end{align}
It has an interpretation similar to that in eq.~\eqref{eq:AR_ann}, as sketched in Fig.~\ref{fig:Eddies_AR_diagram}. However, the anisotropy metrics $\mathrm{AR}^\mathrm{shell}$ and $
\mathrm{AR}^\mathrm{mar}$ can be differ quantitatively due to restrictions on ${\bK}^\mathrm{mar}$ explained in Fig.~\ref{fig:AnisotropySpectraSchema_marginal}.

Below, we provide numerical examples to demonstrate how these metrics quantify anisotropy and to highlight their limitations relative to shell-based metrics presented in the main text. 

\subsection{Centroids and anisotropic metric associated with marginal spectra}
\begin{figure}[bhp]
\centering 
\begin{minipage}[b]{1.0\textwidth}  
\centering
\subfigure[Gauss centroid, Gaussian filter]
{\includegraphics[width=0.42\textwidth]{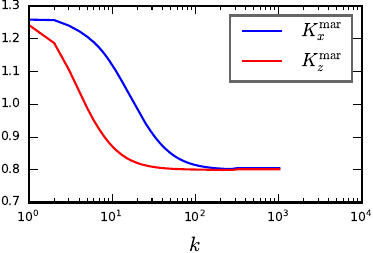} }
\subfigure[Gauss anisotropy metric, Gaussian filter]
{\includegraphics[width=0.42\textwidth]{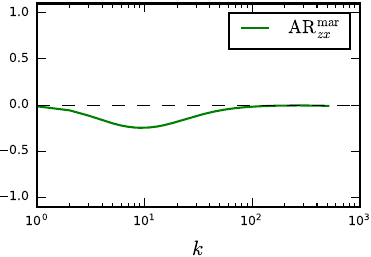} }\\
\subfigure[Gauss centroid, sharp spectral filter]
{\includegraphics[width=0.42\textwidth]{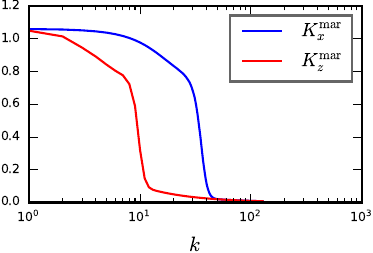} }
\subfigure[Gauss anisotropy metric, sharp spectral filter]
{\includegraphics[width=0.42\textwidth]{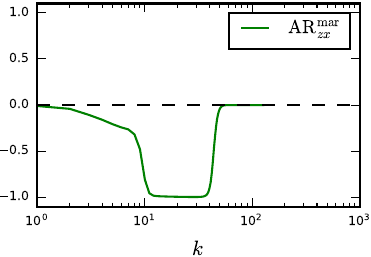} }\\
\subfigure[TG vortex centroid, sharp spectral filter]
{\includegraphics[width=0.42\textwidth]{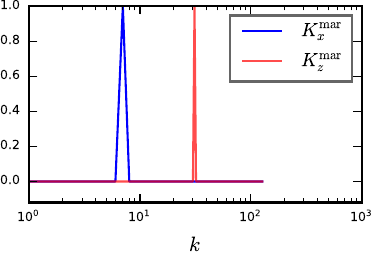} }
\subfigure[TG vortex anisotropy metric, sharp spectral filter]
{\includegraphics[width=0.42\textwidth]{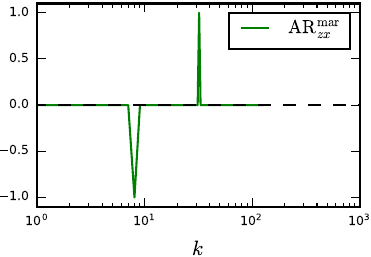} }\\
    \caption{Marginal spectra averaged centroid (eq.~\eqref{eq:integral_K_mar}) and anisotropic metric (eq.~\eqref{eq:AR_mar}) of the Gaussian scalar field using Gaussian filters in panels (a),(b), and of the Gaussian scalar using the sharp spectral filter in panels (c), (d), and of the Taylor-Green vortex using the sharp spectral filter in panels (e),(f).
 \label{fig:marginal_aniso_metrics_manufactured}}
\end{minipage}
\end{figure}

\begin{figure}[bhp]
\centering 
\begin{minipage}[b]{1.0\textwidth}  
\centering
\subfigure[centroid of 3D RT KE, x-z]
{\includegraphics[width=0.42\textwidth]{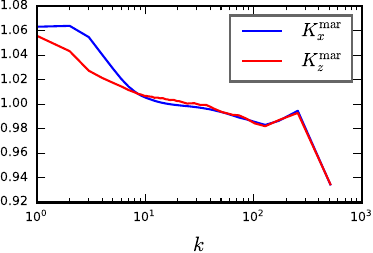} }
\subfigure[anisotropy metric of 3D RT KE, x-z]
{\includegraphics[width=0.42\textwidth]{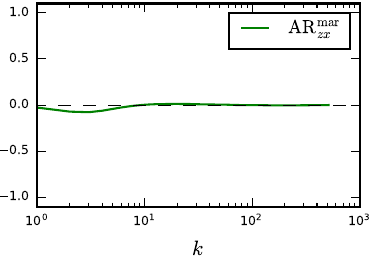} }\\
\subfigure[centroid of 3D RT KE, x-y]
{\includegraphics[width=0.42\textwidth]{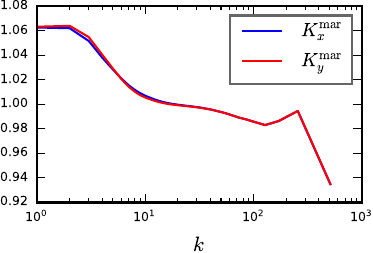} }
\subfigure[anisotropy metric of 3D RT KE, x-y]
{\includegraphics[width=0.42\textwidth]{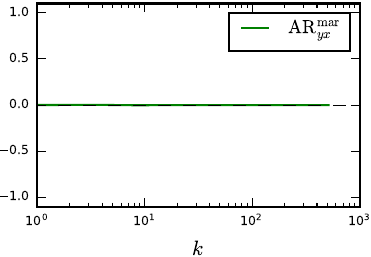} }\\
\subfigure[centroid of 2D RT KE, x-z]
{\includegraphics[width=0.42\textwidth]{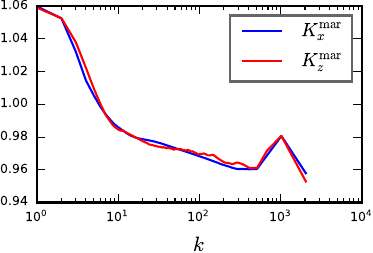} }
\subfigure[anisotropy metric of 2D RT KE, x-z]
{\includegraphics[width=0.42\textwidth]{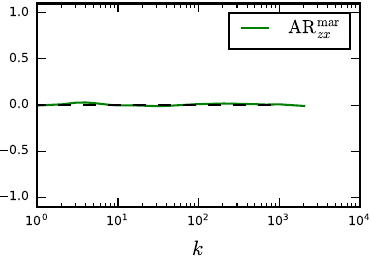} }\\
    \caption{Marginal spectra based centroid (eq.~\eqref{eq:integral_K_mar}) and anisotropic metric (eq.~\eqref{eq:AR_mar}) of the 3D Rayleigh-Taylor kinetic energy along the $x$-$z$ directions in panels (a),(b), and of 3D RT KE along $x$-$y$ directions in panels (c), (d), and of kinetic energy of 2D RT in panels (e),(f).
 \label{fig:marginal_aniso_metrics_RT}}
\end{minipage}
\end{figure}

\lb{sec:AppMargSpec}
The anisotropy metric based on the marginal spectra is shown in Fig.~\ref{fig:marginal_aniso_metrics_manufactured} for the two illustrative examples, and in Fig.~\ref{fig:marginal_aniso_metrics_RT} for the 2D and 3D Rayleigh-Taylor fields.

Fig.~\ref{fig:marginal_aniso_metrics_manufactured}(a) plots $K^{\textrm{mar}}_x(k)$ and $K^{\textrm{mar}}_z(k)$ from eq.~\eqref{eq:integral_K_mar} for the Gaussian data using Gaussian filters. Consistent with their $\bK^{\textrm{shell}}$ analogue shown in fig.~\ref{fig:anisotropic_gaussian}c, it shows that $K^{\textrm{mar}}_z \le K^{\textrm{mar}}_x$ over the entire range of $k$, which implies that the field is elongated along the $z$-direction at all scales. We see that $K^{\textrm{mar}}_z \approx K^{\textrm{mar}}_x$ in the limits of $k\to 0$ and $k\to \infty$. The anisotropy metric $\mathrm{AR}_{zx}^\mathrm{mar}(k)$ in panel (b) from eq.~\eqref{eq:AR_mar} reflects this behavior and is qualitatively consistent with its analogue $\mathrm{AR}_{zx}^\mathrm{shell}(k)$ in fig.~\ref{fig:anisotropic_gaussian}d, although with seemingly lower sensitivity. This is not surprising given that $\bK^{\textrm{mar}}(k)$ is derived from a highly reduced spectral representation as discussed in section~\ref{sec:MarginalSpectra} (see also Fig.~\ref{fig:AnisotropySpectraSchema_shell}). 
The sharp spectral filtering results shown in panels (c) and (d) are similar to the Gaussian kernel results in panels (a) and (b).

Fig.~\ref{fig:marginal_aniso_metrics_manufactured} (e) and (f) show for the Taylor-Green data the centroid $\bK^{\textrm{mar}}$ based on marginal spectra, which exhibits a spike in $K_x^{\textrm{mar}}$ at $k=7$ and a spike in $K_z^{\textrm{mar}}$ at $k=31$. This is because $\OL{E}_x^{\textrm{mar}}$ and $\OL{E}_z^{\textrm{mar}}$ spike at those respective wavenumbers. However, they highlight a limitation of marginal spectra we mentioned; since $\OL{E}_x^{\textrm{mar}}$ and $\OL{E}_z^{\textrm{mar}}$ have to yield the same total energy when integrated, if $\OL{E}_x^{\textrm{mar}}(k_1) > \OL{E}_z^{\textrm{mar}}(k_1)$ at $k_1$, then we necessarily have $\OL{E}_x^{\textrm{mar}}(k_2) < \OL{E}_z^{\textrm{mar}}(k_2)$ at another wavenumber $k_2$.
This is also expressed by the anisotropy metric $\mathrm{AR}_{ij}^\mathrm{mar}(k)$ in panel (f), where $\textrm{AR}^{\textrm{mar}}_{zx}(7)=-1$ and $\textrm{AR}^{\textrm{mar}}_{zx}(31)=1$.

The marginal spectra based centroids and anisotropy metrics for the 2D and 3D RT data are shown in fig.~\ref{fig:marginal_aniso_metrics_RT}. For the shape anisotropy along the $x$-$z$ directions for 3D RT, metrics based on marginal spectra in Fig.~\ref{fig:marginal_aniso_metrics_RT}(a),(b) show qualitatively consistent results as the shell-based results in fig.~\ref{fig:3D_RT_KE_xz}(c),(d), albeit with a much weaker anisotropic signal. This is consistent with our earlier remarks about their lower sensitivity to anisotropy (Fig.~\ref{fig:AnisotropySpectraSchema_shell}). After all, $\OL{E}^{\textrm{mar}}$ has significantly less scale information than $\OL{E}^{2\textrm{D}}$. 

The marginal spectra based shape anisotropy for 3D RT along the $x$-$y$ direction in fig.~\ref{fig:marginal_aniso_metrics_RT}(d) is zero, indicating the flow is isotropy among the horizontal directions as expected. For 2D RT data in panels (e), (f), metrics based on
marginal spectra fail to capture the small-scale anisotropy compared to the shell-based results in fig.~\ref{fig:2D_RT_KE_xz}(c),(d), due to their lower
sensitivity discussed in the main text.

\subsection{Local measurement of shape anisotropy}
\begin{figure}[bhp]
\centering 
\begin{minipage}[b]{1.0\textwidth}  
\centering
\subfigure[lower third of the domain]
{\includegraphics[width=0.32\textwidth]{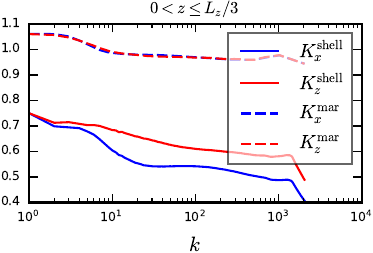} }
\subfigure[middle third of the domain]
{\includegraphics[width=0.32\textwidth]{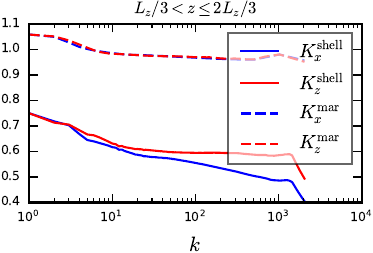} }
\subfigure[upper third of the domain]
{\includegraphics[width=0.32\textwidth]{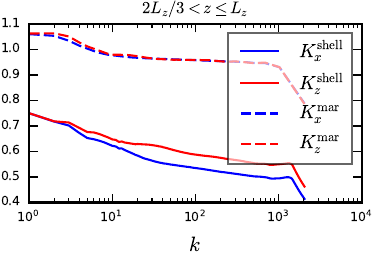} }\\
\subfigure[lower third of the domain]
{\includegraphics[width=0.32\textwidth]{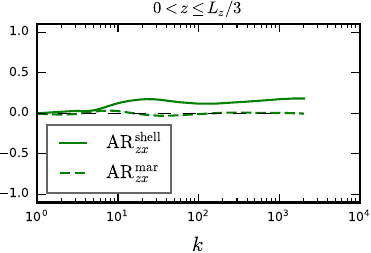} }
\subfigure[middle third of the domain]
{\includegraphics[width=0.32\textwidth]{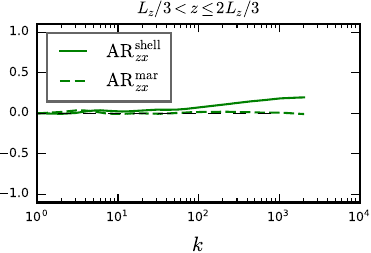} }
\subfigure[upper third of the domain]
{\includegraphics[width=0.32\textwidth]{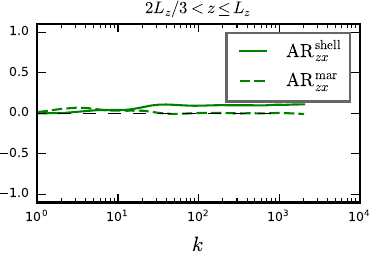} }
    \caption{The centroids (a)-(c) and anisotropy metrics (d)-(f) of 2D RT associated with different vertical regions within $z\in [0, L_z]$. Panels (a,d),(b,e),(c,f) are associated with regions $0<z\leq L_z/3$, $L_z/3 < z\leq 2L_z/3$, and $2L_z/3<z\leq L_z$, respectively.
 \label{fig:regional_aniso_metrics}}
\end{minipage}
\end{figure}

\lb{sec:AppLocalAnisotropy}
One advantage of our filtering approach is that we can measure the anisotropy associated with local sub-regions of the flow in addition to the whole domain. Fig.~\ref{fig:regional_aniso_metrics} shows the centroids of the 2D RT corresponding to three different sub-regions along vertical direction. The results indicate that the three sub-regions are qualitatively similar but exhibit slight quantitative differences in their anisotropy. We defer further analysis to future work.

\subsection{Isotropy of 3D-RT at the domain-size scales}\lb{sec:App3DRTisotropy}
As we discuss in the main text, at the largest possible scales, coarse-grained fields approach the spatial mean, which is trivially isotropic. Specifically, $\OL\phi_\vell \to \{\phi\}$ as $\vell \to (\infty,\infty)$. The spatial mean, $\{\phi\}$, is uniform and, thus, trivially isotropic. This can be easily seen from the illustrative example of an anisotropic Gaussian in Fig.~\ref{fig:synthetic_data_viz}(a), which is isotropic at scales comparable to the domain size as shown in Fig.~\ref{fig:Gaussian_filtered_Gaussian_viz}. This is why the anisotropic metric $AR^{\textrm{shell}}_{zx}\approx 0$ at $k=1$ in Fig.~\ref{fig:anisotropic_gaussian}(d).

\begin{figure}[h]
\centering
\includegraphics[width=.4\textwidth]{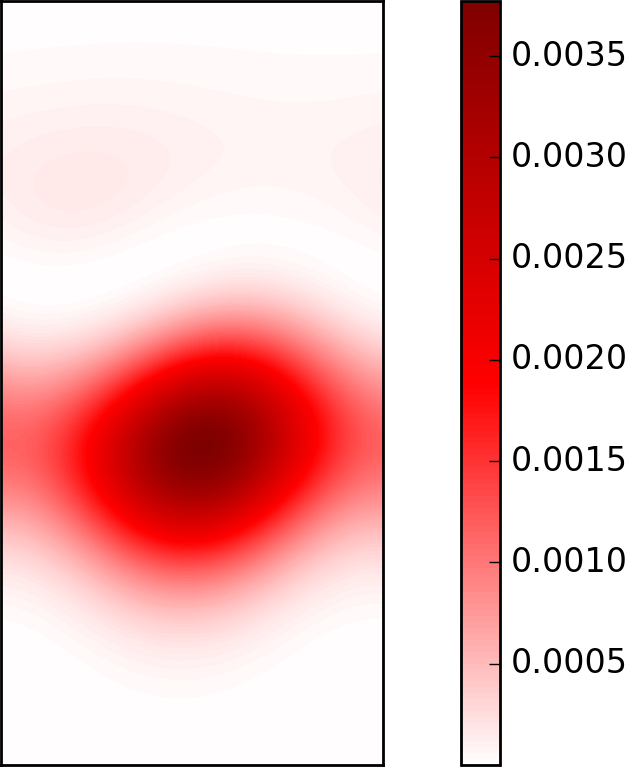}	
	\caption{\footnotesize{Visualization from 3D-RT of coarse KE $\frac{1}{2}\overline{\rho}|\widetilde{\bu}|^2$ filtered at $\ell=L_x$, corresponding to $k=1$.}}
	\label{fig:filtered_KE_viz}
\end{figure}

 For 3D-RT, we have a similar situation in Fig.~\ref{fig:3D_RT_KE_xz}(d) where $AR^\mathrm{shell}\approx 0$ at the horizontal domain scale $k=1$ (\textit{i.e.} $\ell=L$, where $L=3.2$ is the domain size along the $x$-direction). This can be seen from the visualization of coarse KE at $k=1$ in Fig.~\ref{fig:filtered_KE_viz}. Coarse KE is isotropic because the filter width is comparable to the horizontal domain size, such that the coarse-grained fields approache the spatial mean, which is trivially isotropic.  Note that these large scales ($k=1$) where isotropy exists are larger than the size of biggest bubble or spike in 3D-RT seen in Fig.~\ref{fig:RT_viz} and Fig.~\ref{fig:3D_RT_rho_KE_filter_viz}.

\subsection{Insensitivity of our results to filtering kernels}
\begin{figure}[bhp]
\centering 
\begin{minipage}[b]{1.0\textwidth}  
\centering
\subfigure
{\includegraphics[width=0.5\textwidth]{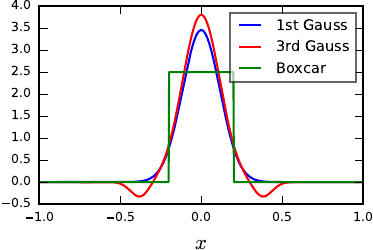} }
    \caption{The Boxcar, first- and third-order Gaussian filtering kernels of width $\ell=L_x/8=0.4$.
 \label{fig:1st_3rd_kernel}}
\end{minipage}
\end{figure}

\lb{sec:AppChangeKernels}
To verify that our results are not sensitive to the particular filtering kernel adopted in the paper, which is a first-order Gaussian \cite{Sadek18}, we repeat our analysis with a Boxcar kernel, and with a third-order modified Gaussian filter. The expression of the third-order kernel $G^{I}_\ell(x)$ with filtering size $\ell$ is
\begin{align}\lb{eq:HighOrderGauss}
G^{I}_\ell(x) = c \,G_\ell(x,\ell)-c' \,G_{\ell'}(x-x_0) - c' \,G_{\ell'}(x+x_0)
\end{align}
where $x$ is the spatial variable, $G_\ell(x,\ell)=\sqrt{\frac{6}{\pi\ell^2}}e^{-\frac{6x^2}{\ell^2}}$ is the first-order Gaussian kernel as in eq.~\ref{eq:GaussKernel}, $c=1.1, c'=0.05, \ell'=0.5\ell, x_0=\sqrt{\frac{43}{48}}\ell$ are parameters to satisfy the third-order kernel constraints $\int G^{I}_\ell \,dx =0$ and $\int x^3 G^{I}_\ell \,dx=0$. This higher-order kernel is, in some sense, more similar to a truncated Dirichlet kernel \cite{spyropoulos1994evaluation} than a Guassian. The shapes of the first- and the third-order Gaussian, as well as the Boxcar, kernels are shown in fig.~\ref{fig:1st_3rd_kernel}.

\begin{figure}[bhp]
\centering 
\begin{minipage}[b]{1.0\textwidth}  
\centering
\subfigure[centroid, first-order Gaussian]
{\includegraphics[width=0.42\textwidth]{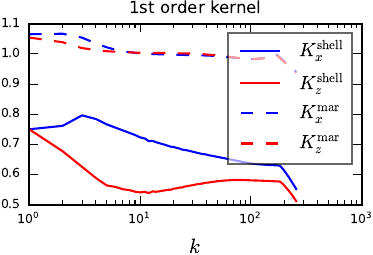} }
\subfigure[anisotropy metric first-order Gaussian]
{\includegraphics[width=0.42\textwidth]{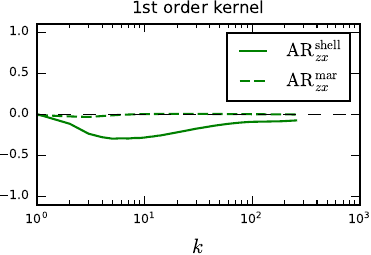} }\\
\subfigure[centroid, boxcar kernel]
{\includegraphics[width=0.42\textwidth]{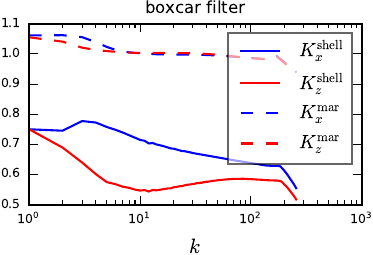} }
\subfigure[anisotropy metric boxcar kernel]
{\includegraphics[width=0.42\textwidth]{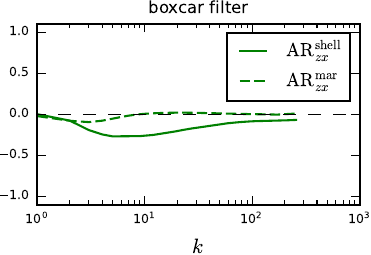} }\\
\subfigure[centroid, third-order Gaussian]
{\includegraphics[width=0.42\textwidth]{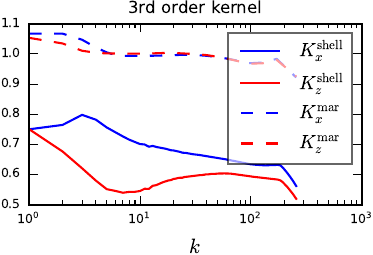} }
\subfigure[anisotropy metric third-order Gaussian]
{\includegraphics[width=0.42\textwidth]{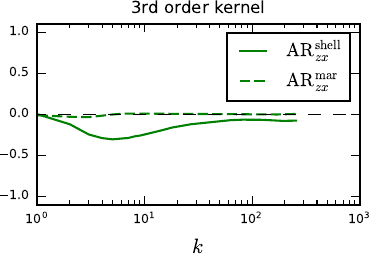} }
    \caption{Centroid (eq.~\eqref{eq:integral_K_mar}) and anisotropic metric (eq.~\eqref{eq:AR_mar}) of the 3D Rayleigh-Taylor kinetic energy within a $x$-$z$ slice. Panels (a),(b) are for first-order kernels adopted in the main paper, panels (c),(d) are the results of top-hat or boxcar filter, while panels (c),(d) are obtained with third-order kernel of eq.~\ref{eq:HighOrderGauss}.
 \label{fig:1st_3rd_anisotropy_metrics}}
\end{minipage}
\end{figure}

\begin{figure}[h]
\centering
\includegraphics[width=.4\textwidth]{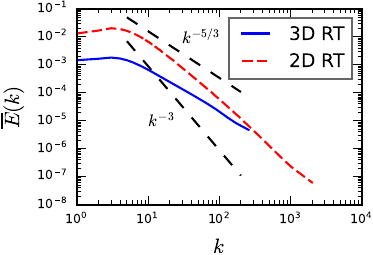}
	\includegraphics[width=.4\textwidth]{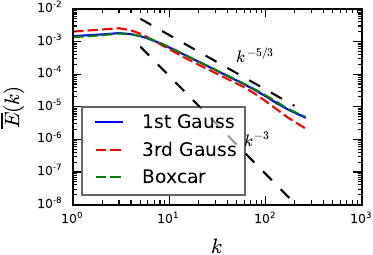}
	\caption{\footnotesize{Left panel: The shell-integrated (or 1D spetrum in eq.~\eqref{eq:FilteringSpectrumVD}) kinetic energy spectra associated with the 2D and 3D RT data used in the paper. Right panel: from 3D-RT data, we show the shell-integrated kinetic energy spectra using three different filtering kernels: first- and third-order Gaussian and a boxcar filter. A $k^{-5/3}$ and a $k^{-3}$ scaling is included in each figure.}}
	\label{fig:KE_spec_three_kernels}
\end{figure}

For convenience, we sample an $x$-$z$ slice from the 3D Rayleigh-Taylor data and perform the anisotropy metrics using the above three kernels. Fig.~\ref{fig:1st_3rd_anisotropy_metrics} and fig.~\ref{fig:KE_spec_three_kernels}(b) show the centroids, the anisotropic metrics, and the shell-averaged 1-D filtering spectra of KE, in which both the qualitative trends and the quantitative values are similar. Thus our shape anisotropy results for RT are not sensitive to the particular filtering kernel we have chosen. 

\clearpage
\subsection{Structure function scalings of 1D fields}
\begin{figure}[bhp]
\centering 
\begin{minipage}[b]{1.0\textwidth}  
\centering
\subfigure[1D field with a (white) spectrum, $E(k)\sim k^0$]
{\includegraphics[width=0.8\textwidth]{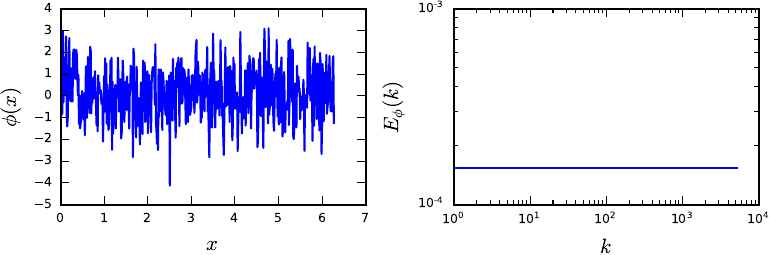} }
\subfigure[1D field with spectrum $E(k)\sim k^{-3}$]
{\includegraphics[width=0.8\textwidth]{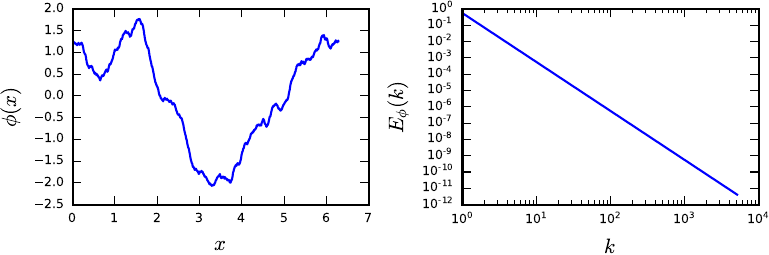} }
    \caption{One-dimensional fields in a periodic domain $[0,2\pi)$ whose Fourier spectra scale as $E(k)\sim k^0$ (top row) and $E(k)\sim k^{-3}$ (bottom row).
 \label{fig:1D_fields_with_scaling}}
\end{minipage}
\end{figure}

\begin{figure}[bhp]
\centering 
\begin{minipage}[b]{1.0\textwidth}  
\centering
\subfigure
{\includegraphics[width=0.45\textwidth]{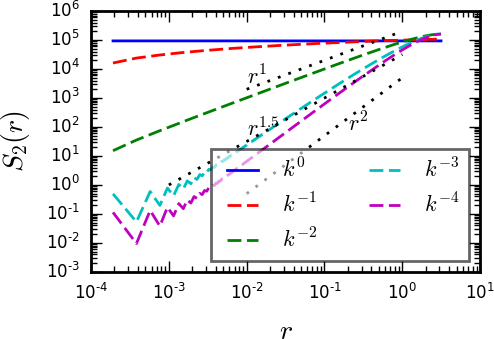} }
\subfigure
{\includegraphics[width=0.45\textwidth]{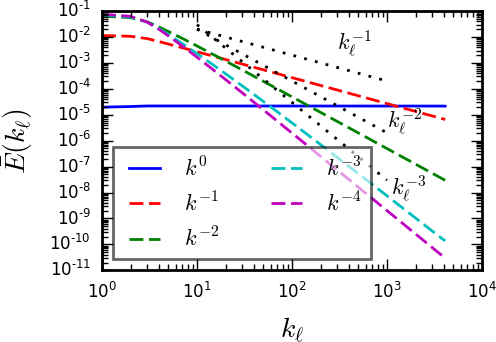} }
    \caption{{\bf Left panel:}  power-law scaling of a 2nd-order structure function, $S_2(r)\sim r^{\alpha-1}$, is related to scaling of the Fourier spectrum $E(k)\sim k^{-\alpha}$ (legend), but only when $1<\alpha<3$. We see that the scaling of $S_2(r)$ is no longer related to that of the Fourier spectrum when $E(k)$ is steeper than $k^{-3}$ (purple) or shallower than $k^{-1}$ (blue). The fields analyzed here are 1D periodic data similar to those shown in Fig.~\ref{fig:1D_fields_with_scaling}.   
    {\bf Right panel:} filtering spectra $\OL{E}(k_\ell)$ using a Gaussian kernel applied to the same data used in the left panel. We see that the filtering spectrum has the same scaling as the Fourier spectrum, $\OL{E}(k)\sim E(k) \sim k^{-\alpha}$, for $\alpha<3$. Specifically, it can correctly capture power-law scaling that is shallower than $k^{-1}$ (blue) but fails for power-law scaling steeper than $k^{-3}$ (purple) since the Gaussian kernel we are using to calculate $\OL{E}(k_\ell)$ is a first-order kernel. It is possible for $\OL{E}(k_\ell)$ to correctly capture  power-laws steeper than $k^{-3}$ by using a higher-order kernel  \cite{Sadek18}.
 \label{fig:1D_fields_structure_func}}
\end{minipage}
\end{figure}

Here, we present the scaling of structure functions associated with periodic 1D field data $\phi(x)$ whose Fourier spectra follows a power-law scaling $E_\phi(k)\sim k^{-\alpha}$, for a few different $\alpha$ values. The field $\phi(x)$ is defined in a periodic domain $x\in [0, 2\pi)$, discretized on a uniform numerical grid of 32768 points. $\phi(x)$ and its associated Fourier spectra are shown in Fig.~\ref{fig:1D_fields_with_scaling}. The corresponding second-order structure function $S_2(r)=\langle (u(x+r)-u(x))^2\rangle$, where the spatial average $\langle \dots\rangle$ is over $x\in [0, 2\pi)$, is shown in Fig.~\ref{fig:1D_fields_structure_func} for 1D fields with Fourier spectra ranging from $k^{-4}$ to $k^0$. Fig.~\ref{fig:1D_fields_structure_func}(a) demonstrates that power-law scaling of a 2nd-order structure function, $S_2(r)\sim r^{\alpha-1}$, is related to scaling of the Fourier spectrum $E(k)\sim k^{-\alpha}$ (legend in Fig.~\ref{fig:1D_fields_structure_func}), but only when $1<\alpha<3$. We see from Fig.~\ref{fig:1D_fields_structure_func}(a) that the scaling of $S_2(r)$ is no longer related to that of the Fourier spectrum when $E(k)$ is steeper than $k^{-3}$ (purple) or shallower than $k^{-1}$ (blue). 

Fig.~\ref{fig:1D_fields_structure_func}(b) shows filtering spectra $\OL{E}(k_\ell)$ using a Gaussian (1st-order) filtering kernel applied to the same data used in panel~\ref{fig:1D_fields_structure_func}(a). We see that the filtering spectrum has the same scaling as the Fourier spectrum, $\OL{E}(k)\sim E(k) \sim k^{-\alpha}$, for $\alpha<3$. Specifically, it can correctly capture power-law scaling that is shallower than $k^{-1}$ (blue) but fails for power-law scaling steeper than $k^{-3}$ (purple) since the Gaussian kernel we are using to calculate $\OL{E}(k_\ell)$ is a 1st-order kernel. It is possible for $\OL{E}(k_\ell)$ to correctly capture  power-laws steeper than $k^{-3}$ by using a higher-order kernel  \cite{Sadek18}.

\clearpage
\subsection{Fourier Spectrum of a Gaussian}\lb{sec:AppGaussianFourierSpectrum}

\begin{figure}[bhp]
\centering 
\begin{minipage}[b]{1.0\textwidth}  
\centering
\subfigure
{\includegraphics[width=0.48\textwidth]{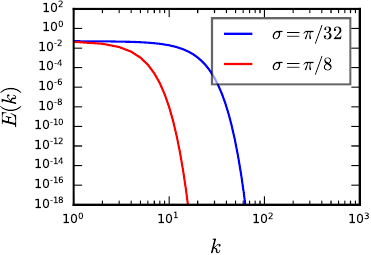} } 
    \caption{Fourier spectra of Gaussians with standard deviations $\sigma=\pi/8$ and $\sigma=\pi/32$
 \label{fig:gaussian_fourier_spec}}
\end{minipage}
\end{figure}

For a 1-dimensional Gaussian function with zero mean and standard deviation $\sigma$, 
\be
\phi(x)=\frac{1}{\sqrt{2\pi}\sigma}e^{-\frac{x^2}{2\sigma^2}},
\ee
defined over the periodic domain $x\in[-\pi, \pi]$, its Fourier transform when $\sigma \ll \pi$ is
\begin{align}\lb{eq:GaussianFFT1D}
\begin{split}
    \widehat{\phi}(k) &= \frac{1}{2\pi}\int_{-\pi}^{\pi} e^{-ikx}\frac{1}{\sqrt{2\pi}\sigma}e^{-\frac{x^2}{2\sigma^2}} dx=\frac{1}{2\pi} e^{-\frac{1}{2}k^2\sigma^2}~.
    \end{split}
\end{align}
The factor $1/2\pi$ in the Fourier transform definition ensures that $\widehat{\phi}(0)$ yields the domain average. Eq.~\eqref{eq:GaussianFFT1D} is an elementary textbook result. Thus, the Fourier spectrum of a Gaussian is
\begin{align}
    E(k) = |\widehat{\phi}|^2(k) + |\widehat{\phi}|^2(-k) = \frac{1}{2\pi^2}e^{-k^2\sigma^2}~.
\end{align}
We drop the factor $\sfrac{1}{2}$ in defining $E(k)$ in the Appendix for convenience. It is straightforward to verify that Parseval's theorem, $\int_0^\infty E(k)dk = \frac{1}{2\pi}\int_{-\pi}^\pi \phi(x)^2 dx$, is satisfied for $\sigma \ll \pi$. Fig.~\ref{fig:gaussian_fourier_spec} compares the Fourier spectra of Gaussians with $\sigma=\pi/8$ and $\sigma=\pi/32$, showing that the narrower Gaussian ($\sigma=\pi/32$) extends to higher $k$.

In our eq.~\eqref{eq:Gaussian} example of the 2D anisotropic Gaussian, we had
\begin{align}
\lb{eq:Gaussian2D}
    \begin{split}
\phi(x,z) = &\frac{1}{\sqrt{2\pi}\sigma_x}\frac{1}{\sqrt{2\pi}\sigma_z}e^{-\frac{x^2}{2\sigma_x^2}-\frac{z^2}{2\sigma_z^2}}~,\\
        \mathrm{where}\hspace{1cm} &(\sigma_x,\sigma_z)=(\pi/32,\pi/8) \hspace{1cm}\mathrm{and}\hspace{1cm}x, z\in [-\pi, \pi]~.
    \end{split}
\end{align}
Its 2-dimensional Fourier transform follows from eq.~\eqref{eq:GaussianFFT1D},
\begin{align}
\lb{eq:FFTGaussian2D}
    \begin{split}
\wh\phi(k_x,k_z) = \frac{1}{(2\pi)^2}e^{-\frac{1}{2}k_x^2\sigma_x^2}~e^{-\frac{1}{2}k_z^2\sigma_z^2}~,
    \end{split}
\end{align}
and its 2D Fourier spectrum is, therefore,
\begin{align}
\lb{eq:GaussianFourierSpectrum2D}
    \begin{split}
E^{2\textrm{D}}(k_x,k_z) = \frac{4}{(2\pi)^4}e^{-k_x^2\sigma_x^2}~e^{-k_z^2\sigma_z^2}~.
    \end{split}
\end{align}


\subsection{Filtering Spectrum of a Gaussian}\lb{sec:AppGaussianFilteringSpectrum}

To obtain the analytical expression for the filtering spectrum of a 1-D Gaussian function $\phi(x)=\frac{1}{\sqrt{2\pi}\sigma} e^{-\frac{x^2}{2\sigma^2}}$, using the filtering kernel $G_\ell(x) = \frac{1}{\sqrt{\pi}\ell}e^{-\frac{x^2}{\ell^2}}$, we first evaluate the filtered quantity $\OL{\phi}_\ell$, then obtain $\{ \OL{\phi}_\ell^2\}$, before calculating the filtering spectrum $\OL{E}(k) = \frac{d}{dk} \{ \OL{\phi}_\ell^2\}$, with $k=L/\ell$ and $L=2\pi$. We now show the details.

The filtered Gaussian $\OL{\phi}_\ell$ is 
\begin{align}\lb{eq:AppFilteredGauss1D}
    \OL{\phi}_\ell = \frac{1}{\sqrt{\pi}\ell}\frac{1}{\sqrt{2\pi}\sigma}\int^{\infty}_{-\infty} \md\xi ~e^{-\frac{(x-\xi)^2}{\ell^2}}e^{-\frac{1}{2}\left(\frac{\xi}{\sigma}\right)^2}~.
\end{align}
Note that in a domain $[-L/2,L/2)$ that is periodic, $\OL{\phi}_\ell$ is calculated in Fourier space. This ensures that eq.~\eqref{eq:AppFilteredGauss1D} is evaluated correctly when $\ell \to L$, since the filtering kernel $e^{-x^2/\ell^2}$ extends beyond $[-L/2,L/2)$ for large $\ell$. Doing so is equivalent to extending the domain periodically in $x$-space beyond its boundaries.

The cumulative spectrum is
\begin{align}
\begin{split}
    \{ \OL{\phi}_\ell^2\} &= \frac{1}{L}\int^{L/2}_{-L/2} \md x~ \frac{1}{\pi\ell^2}\frac{1}{2\pi\sigma^2}\int^{\infty}_{-\infty} \md\xi_1\int^{\infty}_{-\infty} \md\xi_2~ e^{-\frac{(x-\xi_1)^2}{\ell^2}-\frac{(x-\xi_2)^2}{\ell^2}}e^{-\frac{\xi_1^2+\xi_2^2}{2\sigma^2}}\\
    &= \frac{1}{L\sqrt{2\pi}\sqrt{\ell^2+2\sigma^2}} = \frac{1}{L\sqrt{2\pi}\sqrt{L^2k^{-2}+2\sigma^2}}~,
\end{split}
\end{align}
where the second line assumes $\sigma\ll L$ when evaluating integrals involving Gaussians. The assumption is only for the convenience of obtaining an analytical expression.  Finally, the filtering spectrum is
\begin{align}
    \OL{E}(k) = \frac{d}{dk} \{ \OL{\phi}_\ell^2\} = \frac{1}{\sqrt{2\pi}}\frac{ k^{-3}}{L^2(k^{-2}+2\sigma^2/L^2)^{3/2}}~,
\end{align}
which is plotted in Fig.~\ref{fig:gaussian_gaussian_spec} for two different values of $\sigma$. In the limit $k\to \infty$, we see that $\OL{E}(k)\sim k^{-3}$. As discussed in \cite{Sadek18}, this `locking' at $k^{-3}$ scaling overestimates the energy at large $k$ and occurs when calculating the filtering spectrum using a 1st order kernel such as the Gaussian kernel used here. In comparison, the Fourier spectrum $E(k)$ decays faster than $k^{-3}$ as shown in eq.~\eqref{eq:GaussianFFT1D} and Fig.~\ref{fig:gaussian_fourier_spec}. For  the filtering spectrum to be meaningful, one has to use a kernel with at least $p$ vanishing moments when analyzing a function whose Fourier  spectrum decays as $E(k)\sim k^{-\alpha}$ with $\alpha < p+2$. In practice, when Fourier transforms are difficult to perform, one calculates the filtering spectrum using a 1st-order kernel then decides if a higher order kernel is required based on whether or not $\OL{E}(k)$ may be exhibiting a locked $k^{-3}$ scaling. See \cite{Sadek18} for a detailed discussion. 

Note that even with a 1st-order kernel, anisotropy is still detected using filtering spectra. This can be seen from the filtering spectra of the two Gaussians in Fig.~\ref{fig:gaussian_gaussian_spec}, which shows differences in energy at different $k$. However, the differences are less than those in Fig.~\ref{fig:gaussian_fourier_spec} for the Fourier spectra, \textit{i.e.} the level of anisotropy can be underestimated when using a 1st-order kernel.

\begin{figure}[bhp]
\centering 
\begin{minipage}[b]{1.0\textwidth}  
\centering
\subfigure
{\includegraphics[width=0.48\textwidth]{figures/manufactured/Gaussian_gaussian_spec.pdf} } 
    \caption{Filtering spectra of Gaussians with standard deviations $\sigma=\pi/8$ and $\sigma=\pi/32$.
 \label{fig:gaussian_gaussian_spec}}
\end{minipage}
\end{figure}

\subsection{Taylor-Green using the dyadic bands}
\begin{figure}[bhp]
\centering 
\begin{minipage}[b]{1.0\textwidth}  
\centering
\subfigure[~scale-dependent centroid]
{\includegraphics[width=0.42\textwidth]{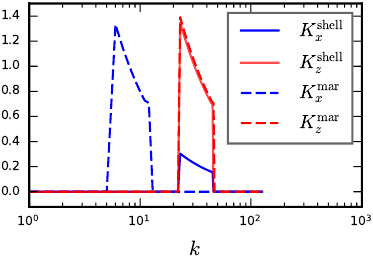} }
\subfigure[~scale-dependent anisotropy metric]
{\includegraphics[width=0.42\textwidth]{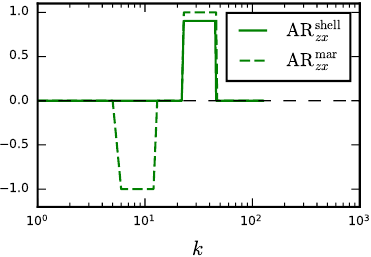} }
    \caption{Similar to Fig.~\ref{fig:anisotropic_TG} for the Taylor-Green velocity (eq.~\eqref{eq:TG}) with a sharp-spectral filter in Fourier space but using the dyadic bands $\bk \in [k/\sqrt{2}, \sqrt{2}k]$. Unlike Fig.~\ref{fig:anisotropic_TG}, no shift of the lines is performed.
 \label{fig:anisotropic_TG_log_band}}
\end{minipage}
\end{figure}

\lb{sec:AppTGLogBand}
Fig.~\ref{fig:anisotropic_TG_log_band} shows centroid and anisotropy metric for the Taylor-Green example using sharp-spectral filter in Fourier space as in Fig.~\ref{fig:anisotropic_TG}  but with dyadic bands $\bk \in [k/\sqrt{2}, \sqrt{2}k]$. 

\subsection{Taylor-Green using Gaussian filter}\lb{sec:AppTGGaussianFilter}

\begin{figure}[bhp]
\centering 
\begin{minipage}[b]{1.0\textwidth}  
\centering
\subfigure[~cumulative spectrum $\mE(k_x,k_z)$]
{\includegraphics[width=0.48\textwidth]{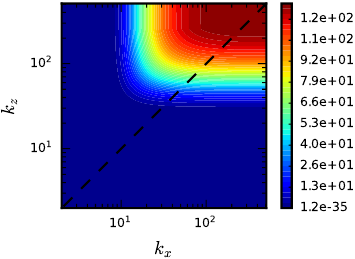} } 
\subfigure[~2D filtering spectrum $\OL{E}^{2\textrm{D}}(k_x,k_z)$]
{\includegraphics[width=0.48\textwidth]{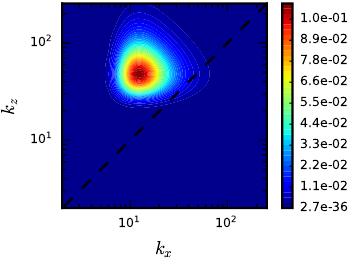}} \\
\subfigure[~scale-dependent centroid]
{\includegraphics[width=0.42\textwidth]{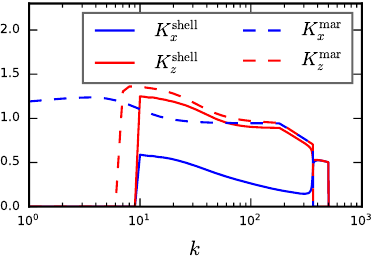} }
\subfigure[~scale-dependent anisotropy metric]
{\includegraphics[width=0.42\textwidth]{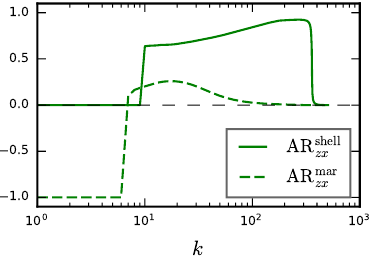} }
    \caption{Similar to Fig.~\ref{fig:anisotropic_TG} for the Taylor-Green velocity field but using the Gaussian filtering kernel. 
    Panels (c) and (d) use bands $k<|\bk|\leq k+1$ as in Fig.~\ref{fig:anisotropic_TG}. 
 \label{fig:anisotropic_TG_Gaussian}}
\end{minipage}
\end{figure}

Fig.~\ref{fig:anisotropic_TG_Gaussian} shows filtering spectra for the Taylor-Green example using a Gaussian kernel. In defining filtering wavenumber, $k=L/\ell$, we use the domain size $L=2\pi$.

\clearpage
\bibliographystyle{unsrt}
\bibliography{RT_citation,Compressible_citation,Turbulence_citation,Numerical_citation}

\end{document}